\documentclass{article}

\usepackage{arxiv}

\usepackage[utf8]{inputenc} 
\usepackage[T1]{fontenc}    
\usepackage{hyperref}       
\usepackage{url}            
\usepackage{booktabs}       
\usepackage{amsfonts}       
\usepackage{nicefrac}       
\usepackage{microtype}      
\usepackage{lipsum}		
\usepackage{graphicx}
\usepackage{doi}

\usepackage{amsmath,amsfonts}
\usepackage{textcomp}
\usepackage{url}
\usepackage{graphicx}
\usepackage{multirow}
\usepackage{subfigure}
\usepackage{diagbox}
\usepackage{scalerel}
\usepackage{hyperref}

\newcommand{\lonet}{LoNet }
\newcommand{\lonetnospace}{LoNet}

\newcommand{\glonet}{GloNet }
\newcommand{\glonetnospace}{GloNet}

\newcommand{\munet}{MuNet }
\newcommand{\munetnospace}{MuNet}

\newcommand{\newnetworkname}{DeepGraviLens }
\newcommand{\newnetworknamenospace}{DeepGraviLens}

\usepackage[normalem]{ulem}
\useunder{\uline}{\ul}{}

\usepackage[autostyle, english = american]{csquotes}
\MakeOuterQuote{"}

\title{\newnetworknamenospace: a Multi-Modal Architecture for Classifying Gravitational Lensing Data}

\author{ \href{https://orcid.org/0000-0001-7906-4987}{\includegraphics[scale=0.06]{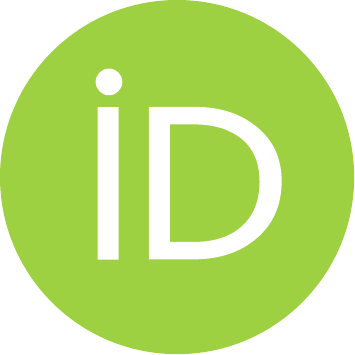}\hspace{1mm}Nicolò Oreste Pinciroli Vago} \\
	Department of Electronics, Information and Bioengineering\\
	Politecnico di Milano\\
	Via Giuseppe Ponzio, 34 \\
	\texttt{nicolooreste.pinciroli@polimi.it} \\
	\And
	\href{https://orcid.org/0000-0002-6945-2625}{\includegraphics[scale=0.06]{orcid.pdf}\hspace{1mm}Piero Fraternali} \\
	Department of Electronics, Information and Bioengineering\\
	Politecnico di Milano\\
	Via Giuseppe Ponzio, 34 \\
	\texttt{piero.fraternali@polimi.it}	
}

\date{}

\renewcommand{\headeright}{}
\renewcommand{\undertitle}{}
\renewcommand{\shorttitle}{\newnetworknamenospace: a Multi-Modal Architecture for Classifying Gravitational Lensing Data}

\hypersetup{
pdftitle={\newnetworknamenospace: a Multi-Modal Architecture for Classifying Gravitational Lensing Data},
pdfsubject={},
pdfauthor={N.O. Pinciroli Vago, P. Fraternali},
pdfkeywords={Multi-modal Deep Learning, Fusion, Gravitational Lensing, Time series},
}

\begin{document}
\maketitle

\begin{abstract}
	Gravitational lensing is the relativistic effect generated by massive bodies, which bend the space-time surrounding them. It is a deeply investigated topic in astrophysics and allows validating theoretical relativistic results and studying faint astrophysical objects that would not be visible otherwise. In recent years Machine Learning methods have been applied to support the analysis of the gravitational lensing phenomena by detecting lensing effects in data sets consisting of images associated with brightness variation time series. However, the state-of-art approaches either consider only images and neglect time-series data or achieve relatively low accuracy on the most difficult data sets. This paper introduces \newnetworknamenospace, a novel multi-modal network that classifies spatio-temporal data belonging to one non-lensed system type and three lensed system types. It surpasses the current state of the art accuracy results by $\approx 3\%$ to $\approx 11\%$, depending on the considered data set. Such an improvement will enable the acceleration of the analysis of lensed objects in upcoming astrophysical surveys, which will exploit the petabytes of data collected, e.g., from the Vera C. Rubin Observatory.
\end{abstract}


\section{Introduction}\label{sec1}

In astrophysics, a gravitational lens is a matter distribution (e.g., a black hole) able to bend the trajectory of transiting light, similar to an optical lens. Such apparent distortion is caused by the curvature of the geometry of space-time around the massive body acting as a lens, a phenomenon that forces the light to travel along the geodesics (i.e., the shortest paths in the curved space-time). Strong and weak gravitational lensing focus on the effects produced by particularly massive bodies (e.g., galaxies and black holes), while microlensing addresses the consequences produced by lighter entities (e.g., stars). This research proposes an approach to automatically classify strong gravitational lenses with respect to the lensed objects and to their evolution through time.

Automatically finding and classifying gravitational lenses is a major challenge in astrophysics. As \cite{doi:10.1146/annurev-astro-081309-130924, shaikh2019strong, islam2021strong, jin2020strong} show, gravitational lensing systems can be complex, ubiquitous and hard to detect without computer-aided data processing. The volumes of data gathered by contemporary instruments make manual inspection unfeasible. As an example, the Vera C. Rubin Observatory is expected to collect petabytes of data \cite{Vicedomini2021}.  

Moreover, strong lensing is involved in major astrophysical problems: studying massive bodies that are too faint to be analyzed with current instrumentation; characterizing the geometry, content and kinematics of the universe; and investigating mass distribution in the galaxy formation process \cite{doi:10.1146/annurev-astro-081309-130924}. Discovery is only the first step, yet a fundamental one, in the study of gravitational lenses. Finding evidence of strong gravitational lensing enables the validation and the advancement of existing astrophysical theories, such as the theory of general relativity \cite{shaikh2019strong}, and supports specialized studies aimed at modeling the effects of gravitational lensing on specific entities, such as wormholes \cite{shaikh2019strong}, Simpson-Visser black holes \cite{islam2021strong}, and Einstein-Gauss-Bonnet black holes \cite{jin2020strong}. 

The gravitational lenses discovery task takes as input spatiotemporal observations consisting of images and time series and associates each observation with one class (e.g., "Lens", "No lens", "Lensed galaxy"...).
Images are obtained from specific regions of the electromagnetic field (e.g., visible and infrared \cite{dahle2002weak}, ultra-violet \cite{quider2009ultraviolet}, and green, red, and near-infrared \cite{morgan2022deepzipper}), depending on the specific experiment. Time series are also collected in specific electromagnetic field regions. They typically describe brightness variation through time (e.g. \cite{morgan2022deepzipper, vakulik2006observational}), and their sampling frequency depends on the technological constraints of the acquisition instrument. In general, they can be multivariate time series \cite{park2021inferring, parkstrongly}. Observations can be either real (i.e., collected by actual instruments) or simulated (i.e., generated by a software system that replicates the characteristics of real instruments).

Several gravitational lenses discovery approaches and tools have been introduced in the past. 
Originally, observations were analyzed without the aid of computers \cite{zwicky1937probability}. Even after the advent of computer science, observations were initially processed without automated classification systems \cite{gorenstein1983detection, lawrence1984discovery, tyson1990detection}. More recently, Machine Learning (ML) methods have been exploited. The works \cite{10.1093/mnras/stz1288, teimoorinia2020comparison} use Convolutional Neural Network (CNNs) to classify gravitational lensing images, \cite{marshall2009automated} exploits a Bayesian approach to categorize image data, and \cite{morgan2022deepzipper} applies a multi-modal approach to classify spatio-temporal data in four simulated data sets generated by the \texttt{deeplenstronomy} simulator \cite{morgan2021deeplenstronomy}. In particular, \cite{morgan2022deepzipper} classifies gravitational lensing data by applying a CNN to the image and a Long Short-Term Memory (LSTM) network to the brightness time series and then fusing the outputs of the two branches, achieving a test accuracy ranging from $48.7\%$ to $78.5\%$.

State-of-the-art lens detection systems, however, still present several limitations. Some of them (e.g., \cite{10.1093/mnras/stz1288, teimoorinia2020comparison, marshall2009automated}) rely on images only and neglect time-domain data and thus cannot detect transient phenomena such as supernovae explosions, which are of great importance for estimating the rate of expansion of the Universe \cite{morgan2022deepzipper2}.  
The work \cite{morgan2022deepzipper} considers spatio-temporal data, but the proposed DeepZipper multi-modal (image + time series) multi-class ("no lens", "lensed galaxy", "lensed type-Ia supernova", "lensed core-collapse supernova") classification architecture shows relatively low accuracy on the most challenging simulated data sets. Moreover, the simulated data set, as presented in \cite{morgan2022deepzipper}, contains $\approx$ 4000 samples in each test set, obtained after an eight-fold data augmentation. The unique samples, before augmentation, then, amount to 500, making the number of test samples of some sub-classes (e.g., "SN-Ia") low ($\approx$ 14 samples) and yielding high uncertainty on the test set accuracy results.

The authors of \cite{morgan2022deepzipper} have recently proposed the DeepZipper II architecture   \cite{morgan2022deepzipper2}, which exploits a multi-modal (image + time series) binary  ("lensed supernova" vs "other")  classification architecture  similar to that of \cite{morgan2022deepzipper},  achieving an accuracy of 93\% over a mix or real and simulated data. The work \cite{10.1093/mnras/stac838}, applies to image time series (i.e., sequences of images), but the classifier works only on the observations where a supernovae is known to be present to infer if lensing has occurred or not.

Multi-modal classification architectures have been exploited in many fields other than astrophysics (e.g., remote sensing and medicine) \cite{ramachandram2017deep, gao2020survey, 10.1162/neco--a--01273}. Only a few approaches consider the combination of a single image and one or more time series \cite{fan2022urban, jayachitra2021cognitive, gadiraju2020multimodal}, and most approaches are similar to the architecture proposed in \cite{morgan2022deepzipper, morgan2022deepzipper2}. Other modalities have been also considered (e.g., videos and texts), but such inputs differ from those relevant to   astrophysical observations and thus such architectures do not carry over the gravitational lensing discovery task. Section \ref{sec:related-work} briefly surveys them.

Finally, the evaluation  of an automatic system for gravitational lens classification poses specific challenges due to the very nature of the task. In real astrophysical observations, gravitational lenses, especially lensed supernovae, are extremely rare  and only a few discoveries have already been validated by the scientific community. The extreme scarcity  of ground truth data (i.e., verified discoveries) challenges both training and testing of classification algorithms and motivates the use of simulators for creating synthetic data sets. Such data sets can be used for training, validating and testing a classifier in the usual way.  
However, when it comes to real data, evaluation can only be done a posteriori by submitting the candidate lensing phenomena to the expert judgement for verification.

This paper presents \newnetworknamenospace, a novel architecture for the classification of strong gravitational lensing multi-modal data. The considered classes concern both transient and non-transient phenomena, and this research shows the superiority of \newnetworkname over other spatio-temporal networks not only at finding gravitational lenses, but also at finding gravitationally-lensed supernovae, rare objects of particular interest to the astrophysical community. The contributions can be summarized as follows:

\begin{itemize}
 \item We introduce the architecture of \newnetworknamenospace, which takes in input spatio-temporal data of real or simulated astrophysical observations and produces in output a multi-class single-label classification of each spatio-temporal sample. 
 \newnetworkname exploits three complementary sub-networks trained independently and combines their outputs by means of a SVM final stage. The three sub-networks apply different and complementary ways of combining image and time-series data, taking advantage of both the local and the global features of the input data. 
 \item We evaluate the designed architecture on four simulated data sets formed by $\approx$ 20000 unique examples, split into a train set with $\approx$ 14000 samples (70\% of the data set), a validation set with $\approx$ 3000 samples (15\% of the data set), and a test set with $\approx$ 3000 samples (15\% of the data set). We compare the predictions of \newnetworkname  to the results obtained by the DeepZipper network \cite{morgan2022deepzipper} and by a version of DeepZipper II \cite{morgan2022deepzipper2} extended from 2 to 4 classes. \newnetworkname yields accuracy improvements ranging from $\approx 10\%$ to $\approx 36\%$ with respect to the best version of DeepZipper on each test set and significantly reduces the confusion between similar classes, one of the major issues of gravitational lenses classification.
 \item We have also compared \newnetworkname with STNet \cite{fan2022urban}, a spatio-temporal multi-modal neural network recently proposed in remote sensing applications, with an improvement in accuracy ranging from $\approx 3\%$ to $11\%$.
 \item Finally, we demonstrate that \newnetworkname is able to detect the presence of gravitational lenses, and specifically gravitationally-lensed supernovae, in real Dark Energy Survey (DES) data \cite{10.1117/12.2312113}.
\end{itemize}

The obtained improvements in the classification of lensing phenomena will enable a faster and more accurate characterization of 
future real observations, such as those of the Vera C. Rubin Observatory, and will open the way to the discovery of lensed supernovae, which are among the hardest bodies to detect due to their rarity, scattered spatial distribution and relatively short observable life \cite{ivezic2019lsst, oguri2019strong, 10.1117/12.2312113, goldstein2016find, 10.1093/mnras/stz1516}. 

The rest of this paper is organized as follows: Section \ref{sec:related-work} surveys the related work; Section \ref{sec:methods} describes the data set and the architecture of \newnetworknamenospace; Section \ref{sec:evaluation} describes the adopted evaluation protocol and presents quantitative and qualitative results; finally, Section \ref{sec:conclusions} draws the conclusions and outlines our future work.

\section{Related Work}
\label{sec:related-work}

This section surveys the previous research in the fields of automated gravitational lensing analysis and multi-modal Deep Learning, which are the foundations of this work.

\subsection{Automated Gravitational Lensing Analysis}

\begin{table}[!htbp]
\label{tab:relworkastro}
\caption{This table summarizes the main approaches for finding gravitational lenses using data-driven techniques. In the "Metric" column, "*" indicates that the metric was computed on real data. The "Real data" column indicates whether the algorithm was tested also on real data, the "Trans." column indicates whether transient phenomena are considered, "LSNe class" indicates whether the "LSNe" class is present in the data set, and "Class. type" is the classification type, which can be either binary (B) or multi-class (M)}
\centering
\resizebox{\columnwidth}{!}{%
\begin{tabular}{@{}lllllllllll@{}}
\toprule
Paper & Year & Algorithm & Survey & Metric & \begin{tabular}[c]{@{}l@{}}Metric\\ result\end{tabular} & \begin{tabular}[c]{@{}l@{}}Real\\ data\end{tabular} & \begin{tabular}[c]{@{}l@{}}LSNe\\ discoveries\end{tabular} & Trans. & \begin{tabular}[c]{@{}l@{}}LSNe\\ class\end{tabular} & \begin{tabular}[c]{@{}l@{}}Class.\\ type\end{tabular} \\ \toprule
\cite{morgan2022deepzipper2} & 2022 & Multi-modal NN & DES & Accuracy & 0.930 & Y & Y & Y & Y & B \\ \midrule
\multirow{2}{*}{\cite{10.1093/mnras/stac838}} & \multirow{2}{*}{2022} & \multirow{2}{*}{Spatiotemporal NN} & YSE & Accuracy & 0.950 & Y & Y & \multirow{2}{*}{Y} & \multirow{2}{*}{Y} & \multirow{2}{*}{B} \\ \cmidrule(lr){4-8}
 &  &  & LSST & Accuracy & 0.990 & N & None &  &  &  \\ \midrule
\cite{savary2022strong} & 2022 & CNN committee & CFIS & Precision* & 0.014 & Y & N/A & N & N & B \\ \midrule
\multirow{4}{*}{\cite{morgan2022deepzipper}} & \multirow{4}{*}{2022} & \multirow{4}{*}{Multi-modal NN} & DES (DES-wide) & Accuracy & 0.487 & N & None & \multirow{4}{*}{Y} & \multirow{4}{*}{Y} & \multirow{4}{*}{M} \\ \cmidrule(lr){4-8}
 &  &  & DES   (DES-deep) & Accuracy & 0.573 & N & None &  &  &  \\ \cmidrule(lr){4-8}
 &  &  & DES   (DESI-DOT) & Accuracy & 0.735 & N & None &  &  &  \\ \cmidrule(lr){4-8}
 &  &  & LSST & Accuracy & 0.785 & N & None &  &  &  \\ \midrule
\cite{stern2021gaia} & 2021 & Tree-based & Gaia & Found lenses* & 14 & Y & N/A & Y & N & B \\ \midrule
\cite{canameras2020holismokes} & 2020 & CNN & Pan-STARRS 3$\pi$ & Accuracy & 0.942 & Y & N/A & N & N & B \\ \midrule
\cite{chao2020lensed} & 2020 & Rule-based & HSC & FPR & 2.30\% & Y & N/A & Y & N & B \\ \midrule
\cite{chan2020survey} & 2020 & HSC & Rule-based & Found lenses* & 6 & Y & N/A & Y & N & B \\ \midrule
\cite{li2020new} & 2020 & CNN & KiDS & FPR & \textless 0.4\% & Y & N/A & N & N & B \\ \midrule
\cite{cheng2020identifying} & 2020 & ConvAE and BGM & Euclid & Accuracy & 0.773 & N & N/A & N & N & B \\ \midrule
\cite{petrillo2019testing} & 2019 & CNN & KiDS & Recall* & 0.750 & Y & N/A & N & N & B \\ \midrule
\cite{delchambre2019gaia} & 2019 & Tree-based & Gaia & AUC* & 0.997 & Y & N/A & N & N & B \\ \midrule
\cite{petrillo2019links} & 2019 & CNN & KiDS & Precision* & 0.025 & Y & N/A & N & N & B \\ \midrule
\cite{khramtsov2019kids} & 2019 & Tree-based & KiDS & Precision* & 0.013 & Y & N/A & N & N & M \\ \midrule
\cite{pearson2018auto} & 2018 & CNN & Various & Accuracy & 0.982 & N & N/A & N & N & B \\ \midrule
\cite{schaefer2018deep} & 2018 & CNN committee & GGSLC & AUC & 0.988 & N & N/A & N & N & B \\ \midrule
\multirow{2}{*}{\cite{hartley2017support}} & \multirow{2}{*}{2017} & \multirow{2}{*}{SVM} & Euclid & AUC & 0.89 & N & \multirow{2}{*}{N/A} & \multirow{2}{*}{N} & \multirow{2}{*}{N} & \multirow{2}{*}{B} \\ \cmidrule(lr){4-7}
 &  &  & KiDS & AUC & 0.95 & Y &  &  &  &  \\ \midrule
\cite{petrillo2017finding} & 2017 & CNN & KiDS & Precision* & 0.029 & Y & N/A & N & N & B \\ \midrule
\cite{marshall2009automated} & 2009 & Bayes & HSC & Completeness & 0.900 & Y & N/A & N & N & M \\ \bottomrule
\end{tabular}%
}
\end{table}

Classifying gravitational lensing phenomena is a challenging task and the subject of many studies. This section concentrates on data-driven techniques, as opposed to the analytical methods that focus on the design of mathematical models capable of explaining the observed data. It considers the specific case of lensed supernovae, as representatives of transient phenomena, as they are particularly interesting for the astrophysics community. Some of the most recent and promising approaches are listed in Table \ref{tab:relworkastro}.

In gravitational lens search, finding lensed supernovae (LSNe) is challenging, as they are rare and fast transient phenomena. The main challenges connected with rarity have been thoroughly analyzed in \cite{savary2022teaching}. A common problem across several lens-finding approaches is the lack of large data sets comprising a sufficient number of real gravitational lens observations. The work \cite{petrillo2019testing}, then, proposes a training set with mock lenses and real non-lensed data, which is a widespread strategy. Several works \cite{morgan2022deepzipper2, canameras2020holismokes, 10.1093/mnras/stac838} also test their trained models on real data and propose some candidate gravitational lenses. 

The second major challenge is considering the transient nature of supernovae. The explosion of a supernova leads to a peak in its brightness, which first increases and can then decline at a slower rate in a few months \cite{hoeflich1996maximum}. The benefits of considering brightness time series in the LSNe case have been illustrated by \cite{morgan2022deepzipper, morgan2022deepzipper2}, and \cite{10.1093/mnras/stac838} uses image time series to consider brightness variability. \cite{morgan2022deepzipper} justifies the extraction of brightness time series from image time series noticing that the differences between images in a series are negligible in 17 representative sub-classes of lensed and non-lensed astrophysical objects. For this reason, their input is formed by a representative image and a normalized brightness time series. The work  \cite{10.1093/mnras/stac838}, instead, uses image time series for finding lensed supernovae and shows promising results on simulated data. However, it considers only two classes: non-lensed supernovae and lensed supernovae, while \cite{morgan2022deepzipper} considers also other astrophysical objects, both lensed and non-lensed, making the input used by \cite{10.1093/mnras/stac838} a particular case of theirs.

The work described in \cite{marshall2009automated} applies a Bayesian approach to classify high-resolution images of non-transient phenomena to reproduce the categorization performed by human experts. However, high-resolution images are not always available and the human classification ("Definitely not a lens", "Possibly a lens", "Probably a lens", "Definitely a lens") is intrinsically imprecise and prone to bias depending on the human classifier.

An alternative to Bayesian methods  \cite{hartley2017support} relies on domain-specific features and separates lensed and non-lensed systems using an SVM, whose output is assessed  by human experts. The classifier obtains, in the best case, an AUC of 0.95 on simulated data, but the presence of manually-defined features makes this approach less general than deep learning methods. In particular, it exploits specific hard-coded characteristics of lenses, such as the prevalence of a specific color, which can hardly generalize to multi-label classification tasks or to scenarios where transient phenomena are relevant. 

Deep learning-based methods rely mostly  on Convolutional Neural Networks (CNNs), as in the binary classifiers illustrated in \cite{10.1093/mnras/stz1288}, \cite{petrillo2019testing} and in \cite{Pourrahmani--2018}, which do not consider time-domain information nor support the fine-grain classification of lensed systems. The work  \cite{canameras2020holismokes} exploits a CNN architecture  and  tests it  also on real data, reporting good results on a binary classification problem that does not focus on LSNe. The authors observe that in their experiments CNN performances relied "heavily on the design of lens simulations and on the choice of negative examples for training, but little on the network architecture". The works  \cite{morgan2022deepzipper, morgan2022deepzipper2}  argue instead that architecture design can lead to great improvements in the results, reporting that multi-modal architectures  outperform single-modality CNNs on transient phenomena data. The work  \cite{pearson2018auto} describes a CNN-based algorithm trained and tested only on simulated data, which  achieves an accuracy of 98\% and  finds the position of the gravitational lens in the input image. However, the classifier is binary and does not consider LSNe. An interesting approach has been proposed in \cite{schaefer2018deep}, which  focuses on the binary classification of simulated data and proposes a committee of networks, yielding an improvement with respect to individual networks.

As an alternative to supervised methods, \cite{cheng2020identifying} defines an unsupervised method for binary classification, which  first uses an autoencoder to denoise the image (reducing its resolution), then applies a second  autoencoder to extract features from the denoised image, and finally exploits a Bayesian Gaussian Mixture (BGM) to cluster the extracted features. This approach, however, requires human intervention  for associating labels to clusters corresponding to the lensed objects. 

Several works focused on finding other gravitationally lensed transient phenomena, such as  quasars. \cite{morgan2022deepzipper} shows that, compared to supernovae,  the brightness of quasars changes in a timescale of several years, because they are not explosive phenomena. For this reason, many studies targeting lensed quasars do not use time series information. \cite{khramtsov2019kids} exploits the image magnitudes in different bands, which is an ad-hoc method that would need adaptation to be applied to the LSNe search. \cite{chao2020lensed} also focuses on finding lensed quasars, but aims at finding quadruply-lensed quasars using an essentially rule-based pipeline. While this method can be effective for the specific application, it should also be modified to tackle more generic and complex cases.

Differently from binary approaches, DeepZipper \cite{morgan2022deepzipper} casts the problem as a multi-class single-label classification task for data sets consisting of images associated with time series of brightness variation. To analyze both images and time-series data, the authors propose a multi-modal network, formed by a CNN and an LSTM, whose outputs are then fused. The resulting system is applied to four simulated data sets  corresponding to different astronomical surveys (DES-wide, LSST-wide, DES-deep, and DESI-DOT). This approach, although relatively simple, achieves relatively good results on all four data sets, with accuracies ranging from $48.7\%$ to $78.5\%$. DeepZipper II \cite{morgan2022deepzipper2},  an evolution of DeepZipper, introduces minor changes to the network, casts the problem as a binary classification task ("LSNe" vs "other") instead of a multi-class  one, and performs testing on a new data set partially based on real data. It reaches an accuracy of $93\%$ on DES data and a false positive rate of $0.02\%$. Three new candidate lensed supernovae found in the DES survey are offered to the astrophysical scientific community for confirmation.

\newnetworknamenospace, similarly to DeepZipper, casts the problem as a multi-class single-label classification task, on the same types of classes and data sets. Compared to previous approaches, it employs more effective unimodal networks and more advanced fusion techniques, which improve the effectiveness in dealing with shared information between the two modalities.

\subsection{Multi-modal Deep Learning and Fusion}

Several phenomena in the most varied disciplines are characterized by heterogeneous data that give complementary information about the subject under investigation. Multi-modal DL has proved its effectiveness in those domains that require the integrated analysis of multiple data types (e.g., images, videos, and time series). The survey  \cite{ramachandram2017deep} overviews the advances and the trends in multi-modal DL until 2017 and documents usage in such areas as medicine \cite{banos2015design, cao2014medical, liang2014integrative}, human-computer interaction \cite{azagra2016multimodal} and autonomous driving \cite{geiger2013vision, maddern20171}. The recent survey \cite{summaira2021recent} discusses several applications combining image and text \cite{liu2020chinese, liu2020image}, video and text \cite{liu2020sibnet, rahman2020semantically}, and text and audio \cite{elias2021parallel, shen2018natural}. Some applications rely on physiological signals for behavioral studies, such as face recognition \cite{cimtay2020cross, li2020multistep, jaiswal2019controlling}.
In the medical field, \cite{shimizu2020artificial} overviews the use of AI in oncology and shows the benefits of multi-modal DL. The work  \cite{10.1007/978-3-319-46723-814} diagnoses cervical dysplasia with the integrated analysis of images and numerical data. \cite{GIBERT2020101873} employs multi-modal DL for classifying malware using textual data from different sources. \cite{velioglu2020detecting} exploits images and texts to detect hate speech in memes. \cite{saito2021select} uses multiple robotic sensors (e.g., cameras, tactile and force sensors) for object manipulation.

From the architecture viewpoint, the processing of heterogeneous inputs can be performed by analyzing the individual data types separately and then fusing the outcome of the different branches to produce an output (late fusion), by stacking the inputs, which are processed together (early fusion), or by introducing fusion at a middle stage (intermediate fusion) \cite{10.1093/bib/bbab569, ramachandram2017deep}. The survey \cite{gao2020survey} overviews DL methods for multi-modal data fusion in general whereas \cite{10.1093/bib/bbab569} focuses on biomedical data fusion. The work 
\cite{summaira2021recent} broadens the comparison beyond DL and contrasts alternative methods employed in multi-modal classification tasks, including SVMs \cite{huang2017fusion}, RNNs \cite{jaiswal2019controlling, wei2020exploiting}, CNNs \cite{hazarika2018icon, oord2018parallel} and even GANs \cite{wei2020multi}. 
The combination of a single image and a time series  has been considered by a few works, mainly in the remote sensing \cite{gomez2015multimodal} and medicine \cite{arioz2022scoping} fields. It is apparently similar to the problem of classifying data formed by a video and a time series \cite{salekin2021multimodal, cai2022deepstroke, pouyanfar2019multimodal}. However, the combination of a single image and a time series, differently from the case of videos, does not require addressing the time-dependent   synchronization, connection and interaction between modalities 
\cite{liang2022foundations}.
Another similar case  is the joint analysis of image and  text. However,  text processing  poses different challenges and adopts different methods with respect to numeric signals \cite{feghali2022overview}. 
Another correlated problem is classifying image time series (i.e., sequences of images), as done in several remote sensing applications (e.g., \cite{suel2021multimodal, pelletier2019temporal, do2019detailed}). This task, addressed also by \cite{10.1093/mnras/stac838} for gravitational lensing data, is best applied when images in the time series vary noticeably. In  gravitational lensing data applications such as the one addressed in this paper, instead, the images in the series have small variations. In such a scenario, the use of time series  is  preferred to the us of image sequences and can be regarded as the extraction of the relevant features from the image sequence \cite{morgan2022deepzipper, morgan2022deepzipper2}.

\subsubsection{Image and time series analysis}

The data considered by \newnetworkname are formed by a single image (the average of the real or simulated observations) and a time series (representing the brightness variation through the observation). Table \ref{tab:info-relwork} summarizes the most representative works based on the combination of a single image and a time series.

\begin{table}[!htbp]
\centering
\caption{This table summarizes representative approaches based on the combination of an image and a time series}
\begin{tabular}{@{}lll@{}}
\toprule
Paper & Year & Field of application \\ \midrule
\cite{ko2022improving} & 2022 & Remote sensing \\
\cite{jacome2022multimodal} & 2022 & Remote sensing \\
\cite{fan2022urban} & 2022 & Remote sensing \\
\cite{manocha2022novel} & 2022 & Medicine \\
\cite{morgan2022deepzipper} & 2022 & Astrophysics \\
\cite{nishimori2021accessory} & 2021 & Medicine \\
\cite{jayachitra2021cognitive} & 2021 & Medicine \\
\cite{gadiraju2020multimodal} & 2020 & Remote sensing \\
\cite{oramas2018multimodal} & 2018 & Music genre classification \\
\cite{vasquez2018multimodal} & 2018 & Medicine \\ \bottomrule
\end{tabular}
\label{tab:info-relwork}
\end{table}

In the medicine field, \cite{vasquez2018multimodal} focuses on the classification of Parkinson's disease severity. It proposes an architecture based on convolutional neural networks to analyze both time series and image data so that the network focuses on local features \cite{zhao2017convolutional}. The authors  show the advantage of a multi-modal approach with respect to the unimodal ones. The work  \cite{pan2023multimodal} considers images, time series, and audio, and proposes a multi-modal approach to classify emotions. The fusion process relies on the computation of RMSE on continuous values predicted by the network, and assigns weights to different modalities based on the errors associated with them. Different from \cite{vasquez2018multimodal}, this method relies on the comparison with GT continuous values (namely, arousal and valence) to determine the weights used during fusion. For this reason, this approach is not extendable to case studies which lack the GT usable for quantifying the prediction errors.
The work \cite{nishimori2021accessory} focuses on diagnosing two heart-related syndromes and proposes the use of ECGs and chest X-rays given in input to a  multi-modal network exploiting CNNs for both images and time series. The works in \cite{manocha2022novel, jayachitra2021cognitive} are two similar approaches that employ multi-modal networks for COVID-19 prediction. Both consider audio signals (for cough, speech, and breathing) and CT scans of the patient's lung. Two networks (one for audio signals, and the other for images) are trained independently. Then, the outcomes are combined with tree-based approaches. In particular, \cite{manocha2022novel} shows that using a decision tree for fusion is more beneficial than using MaxVoting. These approaches are different from the one proposed in \cite{morgan2022deepzipper} because the fusion parameters are learned during a joint training of the two sub-networks.

Representative works in the field of remote sensing have focused on crop yield prediction \cite{jacome2022multimodal}, air pollution prediction \cite{ko2022improving}, crop classification \cite{gadiraju2020multimodal}, and urban informal settlements classification \cite{fan2022urban}. The work  \cite{jacome2022multimodal} proposes an approach for predicting crop yield in Ecuador considering spatio-temporal data. It combines a CNN, for image data, an LSTM, for time series, and a FCNN for late fusion, similarly to the architectures of \cite{morgan2022deepzipper, morgan2022deepzipper2}. The work  \cite{ko2022improving} also focuses on a prediction problem and combines a CNN and an LSTM-based subnetwork. Late fusion is performed by finding the optimal weights associated with each output feature obtained from the unimodal networks. The work  \cite{gadiraju2020multimodal} addresses the problem of crop classification, uses a CNN for the input image and compares different networks (LSTM, CNN, BiLSTM) for the temporal data, showing that the use of a CNN achieves slightly better performances. The final classification step is performed by fusing the unimodal networks decisions using SVM. The work  \cite{fan2022urban} aims at classifying urban informal settlements and proposes a transformer-based approach for fusion.

The combination of images and time series has proven to be beneficial also in the field of music genre classification \cite{oramas2018multimodal}. This work considers the audio signal and the album cover to classify music. The proposed network uses two CNNs to analyze both modalities, similarly to \cite{nishimori2021accessory}, and fuses their feature vectors using a FCNN.

The work in \cite{zhao2023credible} proposes a decision-level fusion approach that leverages the uncertainty associated with each modality, employing a Softplus activation function to quantify uncertainty. This method aims to enhance the credibility of the model's output by considering the uncertainty of each modality, thereby improving the accuracy of the overall results. It has been proposed for generic input modalities, so it can be adapted to the combination of images and time series.

\newnetworkname introduces a novel approach for the classification of images and time series. The proposed architecture exploits three multi-modal networks whose results are assembled using SVM. The three multi-modal networks consider the data in different ways: \lonet exploits intermediate fusion and emphasizes the local features of the image; \glonet applies early fusion and accentuates the global features of both types of input; \munet employs intermediate fusion but extracts both local and global features from the image.

\section{Data Sets and Methods}
\label{sec:methods}

\begin{figure*}
 \centering
 \includegraphics[width=0.8\linewidth]{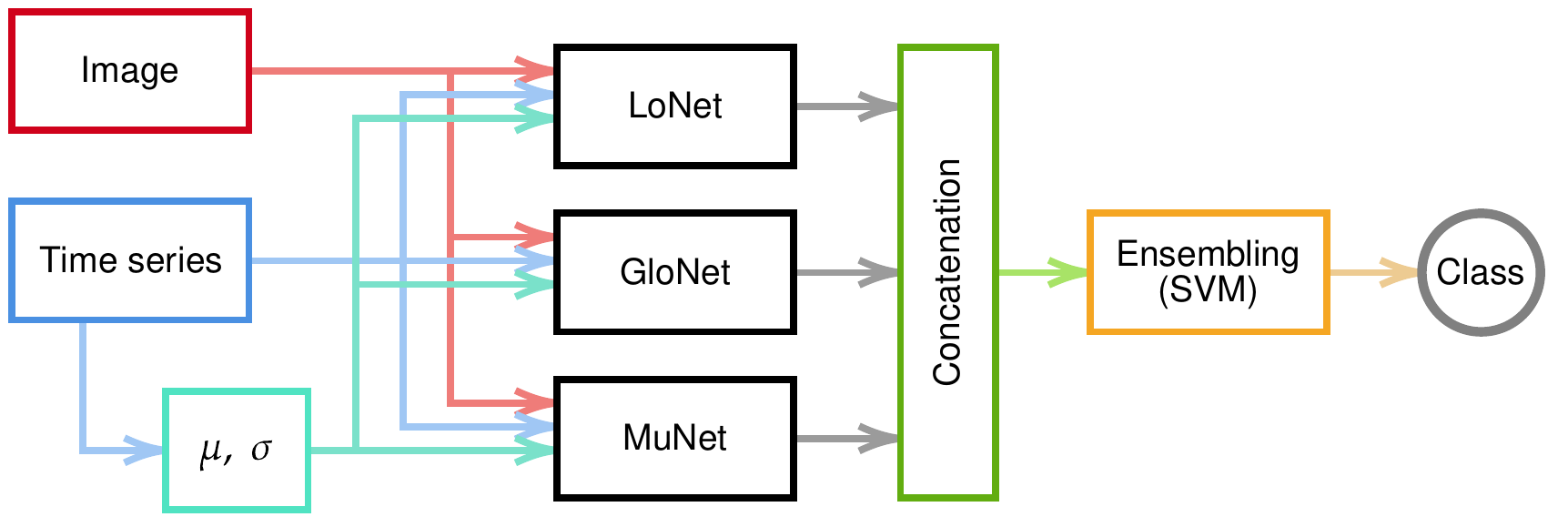}
 \caption {The \newnetworkname pipeline comprises four steps: (1) the inputs are fed into three independent networks (\lonetnospace, \glonetnospace, and \munetnospace); (2) the outputs of the three networks are concatenated; (3) the ReFuse network receives the concatenated outputs and (4) outputs a predicted class}
 \label{fig:pipeline}
\end{figure*}

\subsection{Data sets}
\label{sec:dataset}

An input to the lensed object classification task consists of four images and four brightness variation time series, which together represent an astrophysical observation. One image and one time series are provided for each band of the \textit{griz} photometric system, widely used in CCD cameras \cite{schneider1983ccd}. In this system, the g band is centered on green, the r band is centred on red, the i band is the near-infrared one, and the z band is the infrared one.

Each input is labeled with one of four classes: "No Lens" (no lensed system), "Lens" (Galaxy-Galaxy lensing), "LSNIa" (the lensed object is a Type-Ia supernova), and "LSNCC" (the lensed object is a core-collapse supernova). Section \ref{sec:qualitative-results} shows various examples of input samples and of their classification by \newnetworknamenospace.

Four distinct data sets (DESI-DOT, LSST-wide, DES-wide, and DES-deep) are built via simulation and are used for training and evaluating \newnetworknamenospace.
The details of their construction are similar to the ones presented in \cite{morgan2021deeplenstronomy, birrer2021lenstronomy, morgan2022deepzipper}. Each data set simulates a current or next-generation cosmic survey and is characterized by different specifications of the images and of the associated time series. 

The DESI-DOT data set simulates the observations made by the Dark Energy Camera (DECam) \cite{Flaugher--2015} and mirrors the real observing conditions of the DES wide-field survey reported in \cite{Abbott--2018}. The exposure time, a simulation parameter that affects the image quality (higher is better), was set to 60 seconds. The LSST-wide data set simulates the LSST survey images acquired using the LSSTCam camera \cite{stalder2020rubin}. The simulation parameters were estimated from the conditions of the first year of the survey and the exposure time was set to 30 seconds \cite{phil--marshall--2017--842713}. The DES-wide data set emulates the images from the DECam and uses the real observing conditions from the DES wide-field survey, but the exposure time is 90 seconds. The DES-deep data set also reproduces the images from DECam but its characteristics are simulated according to the DES SN program \cite{Abbott--2019} with the exposure time set to 200 seconds.

Due to the use of the four-bands \textit{griz} photometric system, each image has 4 layers. The image size is $45 \times 45 \times 4$ pixels for all the four data sets. The length of the time series depends on the technical limitations of the simulated instruments. DESI-DOT, LSST-wide, and DES-deep time series contain 14 samples for each band, while DES-wide contains 7 samples for each band.

For each data set, 17 astrophysical systems were defined and grouped into the four classes "No Lens", "Lens", "LSNIa", and "LSNCC" as proposed in \cite{morgan2022deepzipper}. The examples of the four classes were generated randomly: each class covers $\approx 25\%$ of each data set and the distribution of the 17 subsystems is the same in all the data sets. Each data set comprises $\approx 20,000$ elements, split into the train set ($\approx 70\%$), the validation set ($\approx 15\%$), and the test set ($\approx 15\%$).

\subsection{Extraction of statistical quantities}
\label{sec:physical-quantities}
Two statistical quantities (mean $\mu$ and standard deviation $\sigma$) are extracted from the brightness time series and used as inputs. Such derived data have a physical meaning. For example, an empty sky is expected to have approximately the same mean value for the four bands and a high standard deviation (because the fluctuations are random). A non-lensed star is expected to be characterized by a low standard deviation, as the means are approximately constant. Even when they manifest a transient behavior (e.g., the explosion of a supernova), the brightness variation is attenuated by the distance. Lensed bodies instead are expected to have a higher standard deviation, because when they display a transient behavior their brightness is amplified by the lens. The contribution of such derived inputs is quantified in the ablation study described in Section \ref{sec:evaluation}.

\subsection{Overall architecture}

Figure \ref{fig:pipeline} illustrates the multi-stage multi-modal inference pipeline of \newnetworknamenospace. It is formed by three sub-networks (\lonetnospace, \glonetnospace, and \munetnospace), whose outputs are ensembled using SVM. \lonet and \munetnospace, in turn, rely on unimodal sub-networks focusing on local or global features in the images and time series. Table \ref{tab:network-types} summarizes the characteristics of the three networks. \glonet exploits the combination of the image and time-series data, which are merged using early fusion. This approach emphasizes the global features of the multi-modal inputs. \lonet focuses on the local features of the distinct data types: the image and the time series pass through two separate sub-networks and then intermediate fusion is applied. Finally, \munet extracts both local and global features from the image, using an FC sub-network and a CNN in parallel, and then applies intermediate fusion. The next sections present the three proposed multi-modal networks.

\begin{table}[!htbp]
\centering
\caption{The three sub-networks pursue different goals: \glonet emphasizes global features and applies early fusion; \lonet accentuates local features and employs intermediate fusion; \munet extracts both global and local image features}
\label{tab:network-types}
\resizebox{0.5\textwidth}{!}{%
\begin{tabular}{@{}ccc@{}}
\toprule
\multicolumn{1}{l}{\multirow{2}{*}{\textbf{Fusion type}}} & \multicolumn{2}{l}{\textbf{Feature extraction}} \\ \cmidrule(l){2-3} 
\multicolumn{1}{l}{} & \textbf{Global} & \textbf{Local} \\ \midrule
\textbf{Early} & \glonet &  \\ \midrule
\multirow{2}{*}{\textbf{Intermediate}} & \multicolumn{2}{c}{\munet} \\ \cmidrule(l){2-3} 
 &  & \lonet \\ \bottomrule
\end{tabular}%
}
\end{table}

\subsection{\lonetnospace, a network focusing on local features}
\label{sec:net1}

\begin{figure}[ht]
 \centering
 \includegraphics[width=\linewidth]{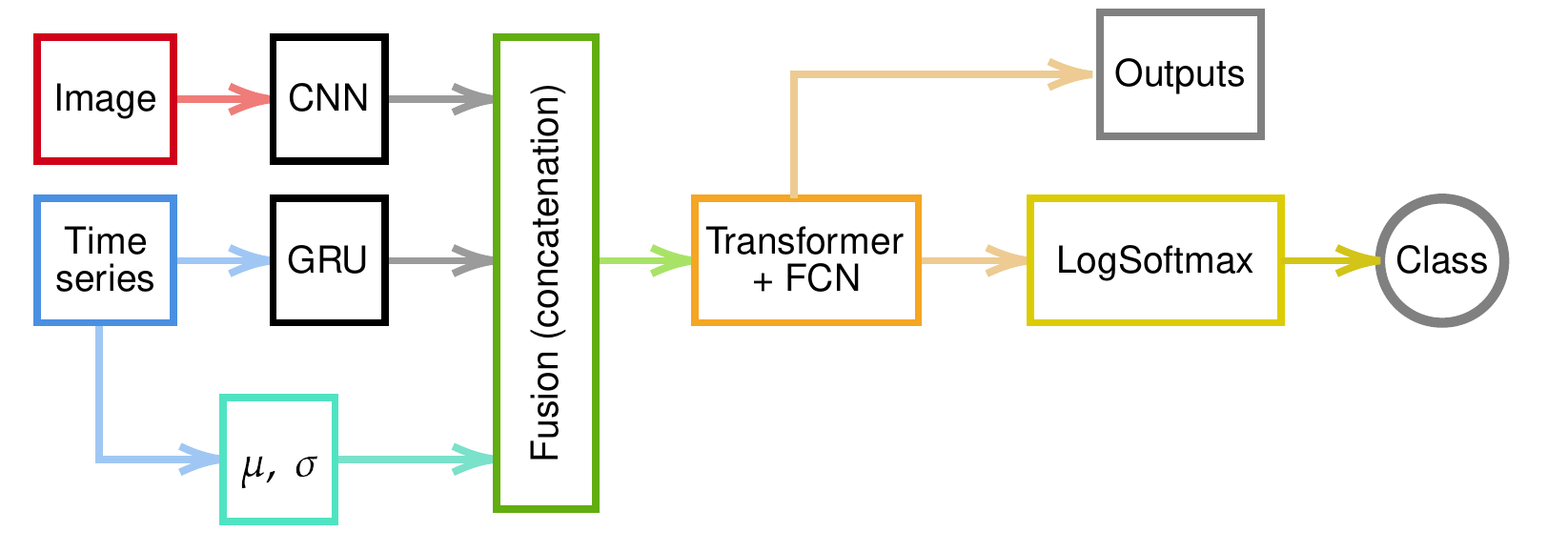}
 \caption{\textbf{\lonet} architecture. The time series is processed by the GRU module and the image by a CNN. The two outputs together with the statistics are fused and fed as input to a final transformer module}
 \label{fig:net1}
\end{figure}

Figure \ref{fig:net1} shows the architecture of the \lonet sub-network and Table \ref{tab:lonet-summary} summarizes its features. It comprises two branches, one for the image (processed through a CNN) and one for the time series (processed by a GRU). 
This structure is similar to the one of ZipperNet \cite{morgan2022deepzipper} but replaces the LSTM \cite{hochreiter1997long} module with a GRU module with a smaller hidden unit size \cite{cho2014learning} and batch normalization. The benefits of GRU over LSTM have been shown in several applications \cite{tang2017memory, chung2014empirical, collins2016capacity, jozefowicz2015empirical, 9221727}. In the considered data sets, the short length of the time series makes GRU advantageous over LSTM because the former has fewer training parameters and thus better generalization abilities.

The use of CNN for extracting features from images privileges the focus on contiguous pixels (i.e., small regions of the image), as shown in several studies \cite{luo2016understanding, pinciroli2021comparing, milani2022proposals}. 

Two feature vectors from the CNN and the GRU, the means and the standard deviations of the time series are concatenated and fed in input to a Transformer, similarly to \cite{fan2022urban}.

\begin{table}[!htbp]
\centering
\caption{Summary of the \lonet neural network architecture showing its layers, output shape, and number of parameters}
\label{tab:lonet-summary}
\resizebox{0.8\textwidth}{!}{%
\begin{tabular}{@{}lll@{}}
\toprule
\textbf{} & \textbf{Output Shape} & \textbf{Parameters \#} \\ \midrule
GRU (features) & [128, 64] & 4,068 \\
CNN (features) & [128, 64] & 2,337,284 \\
Transformer1d: 1-3 & [128, 136] & -- \\
-- TransformerEncoder: 2-15 & [128, 136] & -- \\
---- ModuleList: 3-2 & -- & 2,537,248 \\
Sequential: 1-4 & [128, 32] & -- \\
--       Linear: 2-16 & [128, 64] & 8,768 \\
--       ReLU: 2-17 & [128, 64] & -- \\
--       BatchNorm1d: 2-18 & [128, 64] & 128 \\
--       Dropout: 2-19 & [128, 64] & -- \\
--       Linear: 2-20 & [128, 32] & 2,080 \\
--       ReLU: 2-21 & [128, 32] & -- \\
--       BatchNorm1d: 2-22 & [128, 32] & 64 \\
Sequential: 1-5 & [128, 4] & -- \\
--       Linear: 2-23 & [128, 8] & 264 \\
--       ReLU: 2-24 & [128, 8] & -- \\
--       BatchNorm1d: 2-25 & [128, 8] & 16 \\
--       Dropout: 2-26 & [128, 8] & -- \\
--       Linear: 2-27 & [128, 4] & 36 \\ \midrule
\multicolumn{3}{l}{Total params: 4,889,956} \\
\multicolumn{3}{l}{Trainable params: 4,889,956} \\
\multicolumn{3}{l}{Non-trainable params: 0} \\
\multicolumn{3}{l}{Total mult-adds (G): 3.40} \\ \bottomrule
\end{tabular}%
}
\end{table}

\subsection{\glonetnospace, a network focusing on global features}
\label{sec:net2}

\begin{figure}[ht]
 \centering
 \includegraphics[width=\linewidth]{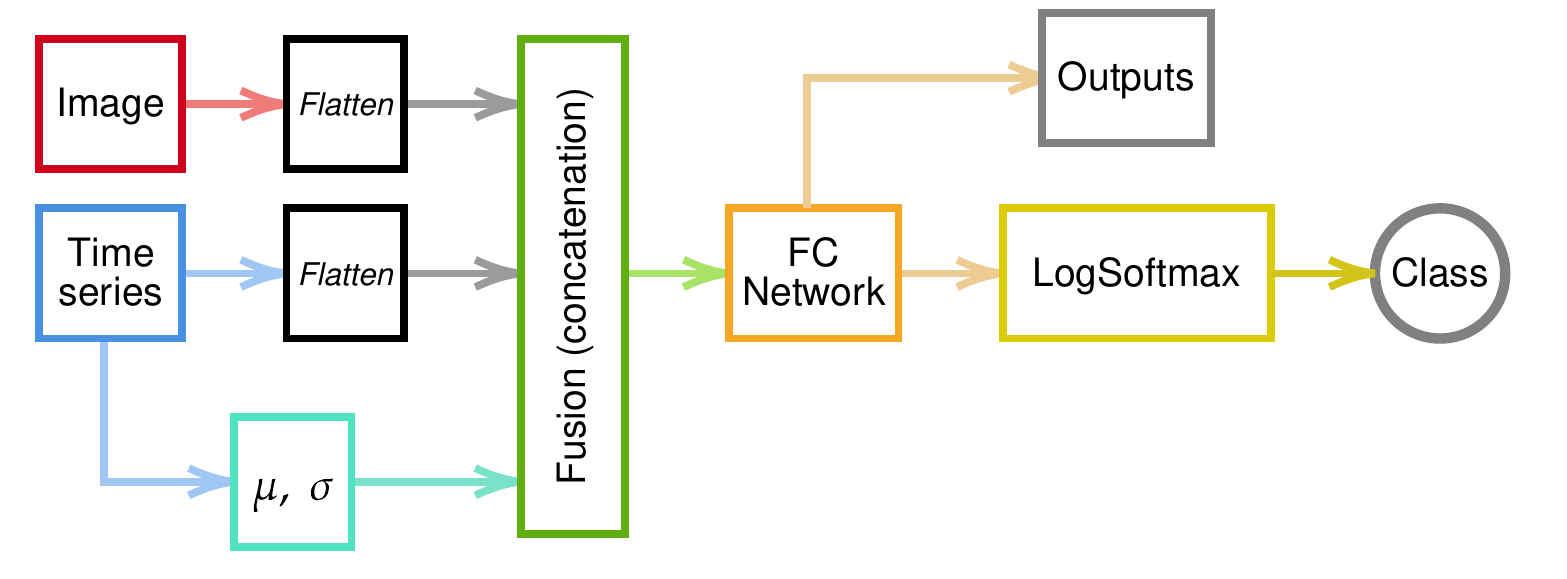}
 \caption{\textbf{\glonet} architecture. The input data are (1) flattened, (2) concatenated, and (3) fed to a FC module}
 \label{fig:net2}
\end{figure}

Figure \ref{fig:net2} shows the architecture of the \glonet sub-network and Table \ref{tab:glonet-summary} summarizes its features. \glonetnospace, differently from \lonetnospace, applies early fusion and relies on a Fully Connected sub-network applied to the flattened inputs. This approach is complementary to the one of \lonetnospace: it combines the original time series and the original image up-front, rather than merging the features derived from their pre-processing by the GRU and CNN modules. Table \ref{tab:glonet-summary} also shows that the number of parameters is higher than in \lonetnospace. Having more parameters allows learning from more complex patterns, which compensates for the absence of convolutional layers.

\begin{table}[!htbp]
\centering
\caption{Summary of the \glonet neural network architecture showing its layers, output shape, and number of parameters. In this case, a time series of 14 steps is considered}
\label{tab:glonet-summary}
\resizebox{0.7\textwidth}{!}{%
\begin{tabular}{@{}lll@{}}
\toprule
 & \textbf{Output Shape} & \textbf{Parameters \#} \\ \midrule
Sequential: 1-1 & [128, 32] & -- \\
--      Linear: 2-1 & [128, 4096] & 33,443,840 \\
--      ReLU: 2-2 & [128, 4096] & -- \\
--      BatchNorm1d: 2-3 & [128, 4096] & 8,192 \\
--      Dropout: 2-4 & [128, 4096] & -- \\
--      Linear: 2-5 & [128, 2048] & 8,390,656 \\
--      ReLU: 2-6 & [128, 2048] & -- \\
--      BatchNorm1d: 2-7 & [128, 2048] & 4,096 \\
--      Dropout: 2-8 & [128, 2048] & -- \\
--      Linear: 2-9 & [128, 1024] & 2,098,176 \\
--      ReLU: 2-10 & [128, 1024] & -- \\
--      BatchNorm1d: 2-11 & [128, 1024] & 2,048 \\
--      Dropout: 2-12 & [128, 1024] & -- \\
--      Linear: 2-13 & [128, 512] & 524,800 \\
--      ReLU: 2-14 & [128, 512] & -- \\
--      BatchNorm1d: 2-15 & [128, 512] & 1,024 \\
--      Dropout: 2-16 & [128, 512] & -- \\
--      Linear: 2-17 & [128, 256] & 131,328 \\
--      ReLU: 2-18 & [128, 256] & -- \\
--      BatchNorm1d: 2-19 & [128, 256] & 512 \\
--      Linear: 2-20 & [128, 128] & 32,896 \\
--      ReLU: 2-21 & [128, 128] & -- \\
--      BatchNorm1d: 2-22 & [128, 128] & 256 \\
--      Dropout: 2-23 & [128, 128] & -- \\
--      Linear: 2-24 & [128, 64] & 8,256 \\
--      ReLU: 2-25 & [128, 64] & -- \\
--      BatchNorm1d: 2-26 & [128, 64] & 128 \\
--      Dropout: 2-27 & [128, 64] & -- \\
--      Linear: 2-28 & [128, 32] & 2,080 \\
Sequential: 1-2 & [128, 4] & -- \\
--      Linear: 2-29 & [128, 8] & 264 \\
--      ReLU: 2-30 & [128, 8] & -- \\
--      Linear: 2-31 & [128, 4] & 36 \\ \midrule
\multicolumn{3}{l}{Total params: 44,648,588}  \\
\multicolumn{3}{l}{Trainable params: 44,648,588} \\
\multicolumn{3}{l}{Non-trainable params: 0} \\
\multicolumn{3}{l}{Total mult-adds (G): 5.72}   \\  \bottomrule
\end{tabular}%
}
\end{table}

\subsection{\munetnospace, a network focusing on local and global features}
\label{sec:net3}

\begin{figure}[ht]
 \centering
 \includegraphics[width=\linewidth]{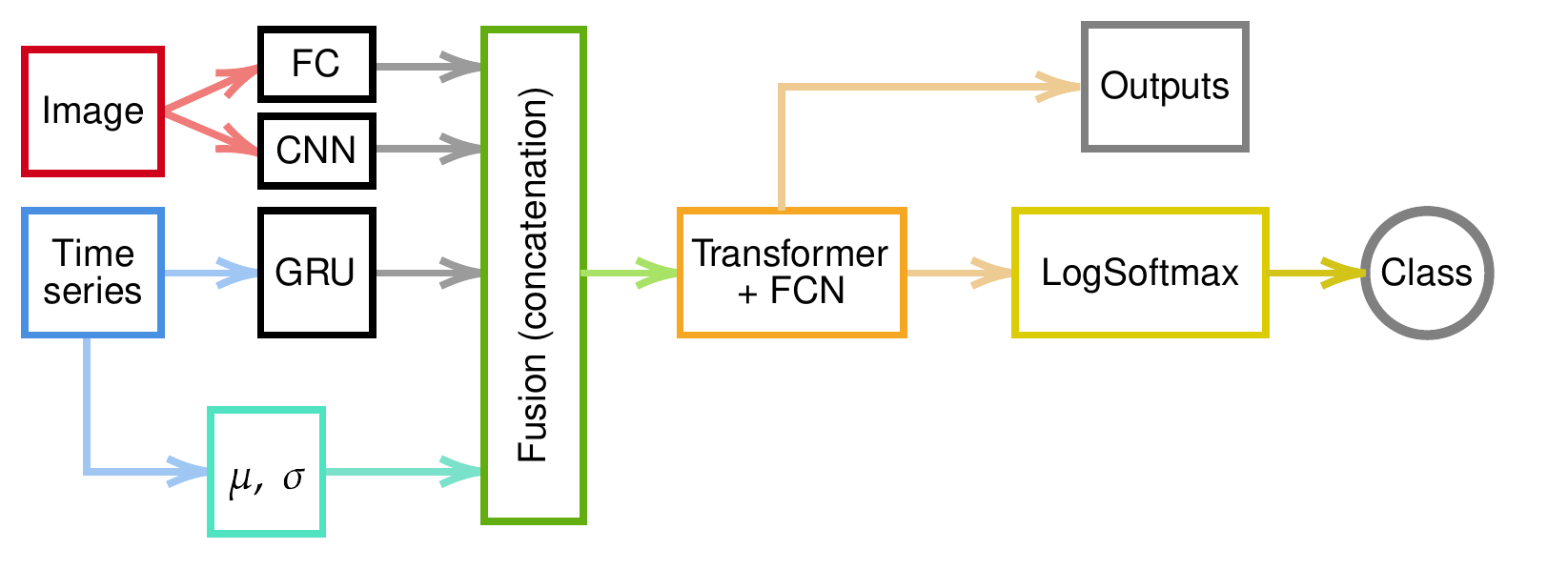}
 \caption{\textbf{\munet} architecture. While \lonet processes the image using only a CNN, \munet employs both a CNN and a FC component}
 \label{fig:net3}
\end{figure}

Figure \ref{fig:net3} shows the architecture of the \munet sub-network and Table \ref{tab:munet-summary} summarizes its features. It processes the image using two parallel branches: a CNN and an FC sub-network. The time series is processed in the same way as in \lonetnospace. 
Compared to \lonetnospace, \munet adds the FC module applied to the image, to extract local and global features simultaneously. The latter may provide a relevant contribution due to the small size of the images. To avoid overfitting, the number of parameters in the FC sub-network is smaller than in \glonetnospace. In total, the number of parameters is similar to the one of \lonetnospace.

\begin{table}[!htbp]
\centering
\caption{Summary of the \munet neural network architecture showing its layers, output shape, and number of parameters}
\label{tab:munet-summary}
\resizebox{0.8\textwidth}{!}{%
\begin{tabular}{@{}lll@{}}
\toprule
\textbf{} & \textbf{Output Shape} & \textbf{Parameters \#} \\ \midrule
FC: 1-1 & [128, 32] & 2,337,284 \\
GRU: 1-2 & [128, 64] & 4,068 \\
CNN: 1-3 & [128, 64] & 2,118,804 \\
Linear: 1-4 & [128, 32] & 2,080 \\
Sequential: 1-5 & [128, 32] & -- \\
     -- Linear: 2-23 & [128, 64] & 8,768 \\
     -- ReLU: 2-24 & [128, 64] & -- \\
     -- BatchNorm1d: 2-25 & [128, 64] & 128 \\
     -- Dropout: 2-26 & [128, 64] & -- \\
     -- Linear: 2-27 & [128, 32] & 2,080 \\
     -- ReLU: 2-28 & [128, 32] & -- \\
     -- BatchNorm1d: 2-29 & [128, 32] & 64 \\
Sequential: 1-6 & [128, 4] & -- \\
     -- Linear: 2-30 & [128, 16] & 528 \\
     -- ReLU: 2-31 & [128, 16] & -- \\
     -- BatchNorm1d: 2-32 & [128, 16] & 32 \\
     -- Dropout: 2-33 & [128, 16] & -- \\
     -- Linear: 2-34 & [128, 8] & 136 \\
     -- ReLU: 2-35 & [128, 8] & -- \\
     -- BatchNorm1d: 2-36 & [128, 8] & 16 \\
     -- Dropout: 2-37 & [128, 8] & -- \\
     -- Linear: 2-38 & [128, 4] & 36 \\ \midrule
\multicolumn{3}{l}{Total params: 4,474,024} \\
\multicolumn{3}{l}{Trainable params: 4,474,024} \\
\multicolumn{3}{l}{Non-trainable params: 0} \\
\multicolumn{3}{l}{Total mult-adds (G): 3.39} \\ \bottomrule
\end{tabular}%
}
\end{table}

\subsection{Ensembling}
\label{sec:ensembling}

\begin{table}[!htbp]
\centering
\caption{Experimental parameters of ensemble methods for aggregating decisions of \lonetnospace, \glonetnospace, and \munetnospace}
\label{tab:ensemble-parametrization}
\resizebox{0.7\textwidth}{!}{%
\begin{tabular}{@{}ll@{}}
\toprule
\textbf{Methods} & \textbf{Parametrization} \\ \midrule
AdaBoost & estimators: 200 \\ \midrule
Random Forest & estimators: [10, 50, 100, 200, 500, 1000, 2000] \\ \midrule
Extra Trees & estimators: 100 \\ \midrule
Fuzzy ranking \cite{manna2021fuzzy} & None \\ \midrule
Average & None \\ \midrule
MLP & \begin{tabular}[c]{@{}l@{}}hidden layer sizes: 100\\ activation: 'relu'\\ solver: 'adam'\\ alpha:  0.0001\end{tabular} \\ \midrule
KNN & neighbours: [2, 4, 6, 8, 16, 32] \\ \midrule
FCNN & Early stop \\ \midrule
Max & None \\ \midrule
SVM & \begin{tabular}[c]{@{}l@{}}C: [$10^{-4}$,   $10^{-3}$, $10^{-2}$, $10^{-1}$, $10^{0}$, $10^{1}$]\\ kernel: ['poly',  'linear', 'rbf', 'sigmoid']\end{tabular} \\ \bottomrule
\end{tabular}%
}
\end{table}

The three multi-modal networks introduced in this study extract distinct information from the data, emphasizing local features, global features, or a combination of both. To fully leverage the complementary information provided by these networks, ensemble methods can be employed. Table \ref{tab:ensemble-parametrization} details the ensemble methods used in this study and their associated experimental parameters. For each parameter combination of every method, accuracy is computed on both the train and validation sets. The best parameter combination is then selected based on the highest validation set result, and the accuracy is finally computed on the test set.
Moreover, an ablation study is conducted to assess the performance of the best ensemble method when using only two out of the three networks.

\subsection{Training}
\label{sec:training}

The training process of \newnetworkname is divided into two stages. In the first step, \lonetnospace, \glonetnospace, and \munet are trained separately, using the same inputs. The second stage consists of  training the  SVM, which exploits as inputs the values obtained before the application of the final activation function of the \lonetnospace, \glonetnospace, and \munet sub-networks. \lonetnospace, \glonetnospace, and \munet are trained for a maximum of 500 epochs, and the Early Stopping patience is set to 20 epochs. In both stages, the best model is the one with the highest validation accuracy.

\section{Evaluation}
\label{sec:evaluation}

This section reports the quantitative and qualitative evaluation of \newnetworkname on the data sets introduced in \ref{sec:dataset}.

For each accuracy result, a confidence interval amounting to 1 standard deviation is calculated to take the limited size of the test set into account. C.R. represents the radius of the confidence interval \cite{witten2002data}:

\begin{equation}
C.R. = \sqrt{\frac{a\cdot(1-a)}{n}}
\end{equation}

where $a$ is the mean accuracy (scaled to $[0, 1]$) on the test set and $n$ is the number of samples in the test set. 

\subsection{Quantitative results}
\label{sec:quantitative-results}

This section presents the outcome of the performance analysis of \newnetworkname on the four data sets described in Section \ref{sec:dataset}. For assessing the improvement induced by the proposed architecture, the approach of \cite{morgan2022deepzipper} is used as a baseline, since it is the only research which used a data set with the same classes as ours. Accuracy is used as the performance metrics because the data set is balanced.
In addition, results were compared with other two multi-modal networks using the time and image modalities, presented in Table \ref{tab:results}, and with seven unimodal networks, presented in Table \ref{tab:comparison-unimodals}. Both DeepZipper II \cite{morgan2022deepzipper2} and STNet \cite{fan2022urban} have been adapted to use four classes rather than the original two.

Ablation experiments with respect to the sub-networks preceding the final ensembling stage are also performed to verify their contribution. 

\subsubsection{Prediction performance}

\begin{table}[!htbp]
\centering
\caption{\textbf{Accuracy} -- Comparison of the accuracy of \newnetworkname and of the best result obtained using state of the art multi-modal methods. An improvement of $\approx 10\%$ to $\approx 36\%$ is achieved with respect to DeepZipper \cite{morgan2022deepzipper}, the only work using a data set with the same classes as \newnetworknamenospace. When compared to the best result obtained by reproducing state of the art approaches, the improvement ranges between $\approx 3\%$ and $\approx 11\%$}
\label{tab:results}
\resizebox{\textwidth}{!}{%
\begin{tabular}{@{}ccccc@{}}
\toprule
 & \textbf{DESI-DOT} & \textbf{DES-deep} & \textbf{DES-wide} & \textbf{LSST-wide} \\ \midrule
\textbf{DeepZipper} \cite{morgan2022deepzipper} & 77.1 & 58.6 & 51.7 & 74.3 \\
\textbf{DeepZipper II} \cite{morgan2022deepzipper2} & 78.9 & 57.4 & 49.8 & 70.7 \\
\textbf{STNet} \cite{fan2022urban} & 85.1 & 58.4 & 82.5 & 84.3 \\  \midrule
\textbf{Evidential\lonet} (Ours) & 81.6 & 65.4 & 79.9 & 84.5 \\
\textbf{Evidential\munet} (Ours) & 81.1 & 65.6 & 79.5 & 84.2 \\ \midrule
\textbf{\lonet} (Ours) & 87.0 & 67.5 & 85.8 & 87.2 \\
\textbf{\glonet} (Ours) & 77.2 & 62.3 & 76.8 & 76.8 \\
\textbf{\munet} (Ours) & 87.9 & 67.9 & 86.5 & 88.5 \\ \midrule
\textbf{\newnetworknamenospace} (Ours) & {\ul \textbf{88.7}} & {\ul \textbf{69.6}} & {\ul \textbf{87.7}} & {\ul \textbf{88.8}} \\
\textbf{Improvement} & 3.6 & 11.0 & 5.2 & 4.5 \\ \bottomrule
\end{tabular}%
}
\end{table}

\begin{table}[!htbp]
\centering
\caption{\textbf{Comparison of the unimodal networks and \newnetworknamenospace} -- The table shows the performance of different unimodal networks on image and time modalities, used in Deep Zipper, STNet, and \newnetworknamenospace. The best unimodal results are highlighted in bold, and the proposed network's performance is underlined}
\label{tab:comparison-unimodals}
\resizebox{\textwidth}{!}{%
\begin{tabular}{@{}cccccccc@{}}
\toprule
\textbf{Modality} & \textbf{Unimodal network} & \textbf{Multi-modal networks} & \textbf{DESI-DOT} & \textbf{DES-deep} & \textbf{DES-wide} & \textbf{LSST-wide} & \textbf{Average} \\ \midrule
\multirow{4}{*}{\textbf{Image}} & ResMixer & STNet & \textbf{81.4} & 65.1 & \textbf{82.5} & \textbf{82.1} & \textbf{77.8} \\
 & CNN (Ours) & \lonetnospace, \munet & 78.4 & \textbf{65.9} & 79.7 & 81.4 & 76.4 \\
 & CNN (DZ) & DeepZipper & 74.3 & 61.9 & 70.0 & 74.3 & 70.1 \\
 & FCNN (Ours) & \munet & 67.7 & 57.7 & 62.3 & 59.3 & 61.8 \\ \midrule
\multirow{3}{*}{\textbf{Time}} & GRU (DZ) & DeepZipper & \textbf{70.9} & 28.5 & \textbf{39.1} & \textbf{67.0} & \textbf{51.4} \\
 & GRU (Ours) & \lonetnospace, \munet & 69.2 & \textbf{32.4} & 38.9 & 61.0 & 50.4 \\
 & PDNet & STNet & 63.8 & 28.5 & 32.9 & 60.9 & 46.5 \\ \midrule
\textbf{Multi-modal} & \multicolumn{2}{c}{\textbf{\newnetworknamenospace} (Ours)} & {\ul \textbf{88.7}} & {\ul \textbf{69.6}} & {\ul \textbf{87.7}} & {\ul \textbf{88.8}} & {\ul \textbf{83.7}} \\
\bottomrule
\end{tabular}%
}
\end{table}

Table \ref{tab:results} presents the accuracy results on the four considered test sets. The test set accuracy is similar for the DESI-DOT, LSST-wide and DES-deep data sets and decreases for the more complex DES-wide data set. In all cases, the accuracy shows an improvement with respect to both the DeepZipper baseline and the best method in the state of the art. Such improvement is observed not only in the case of \newnetworknamenospace, but also for \lonet and    \glonetnospace, making them viable alternatives to state-of-the-art approaches. Moreover, the performance of \glonetnospace, a simple network, are similar to the ones of DeepZipper and DeepZipper II.

In addition to  \lonet and \munetnospace, the networks Evidential\lonet and Evidential\munet  were also implemented and tested. These networks exploit the evidence-based late fusion approach  proposed in \cite{zhao2023credible}, which dynamically weights the contribution of each modality based on the degree of uncertainty associated with its predictions. Our experiments show that the proposed intermediate fusion approach  outperforms the evidence-based fusion approach, with an average improvement of $\approx 4.5\%$.

\begin{figure*}
\centering
\subfigure[]{\includegraphics[width=0.45\textwidth]{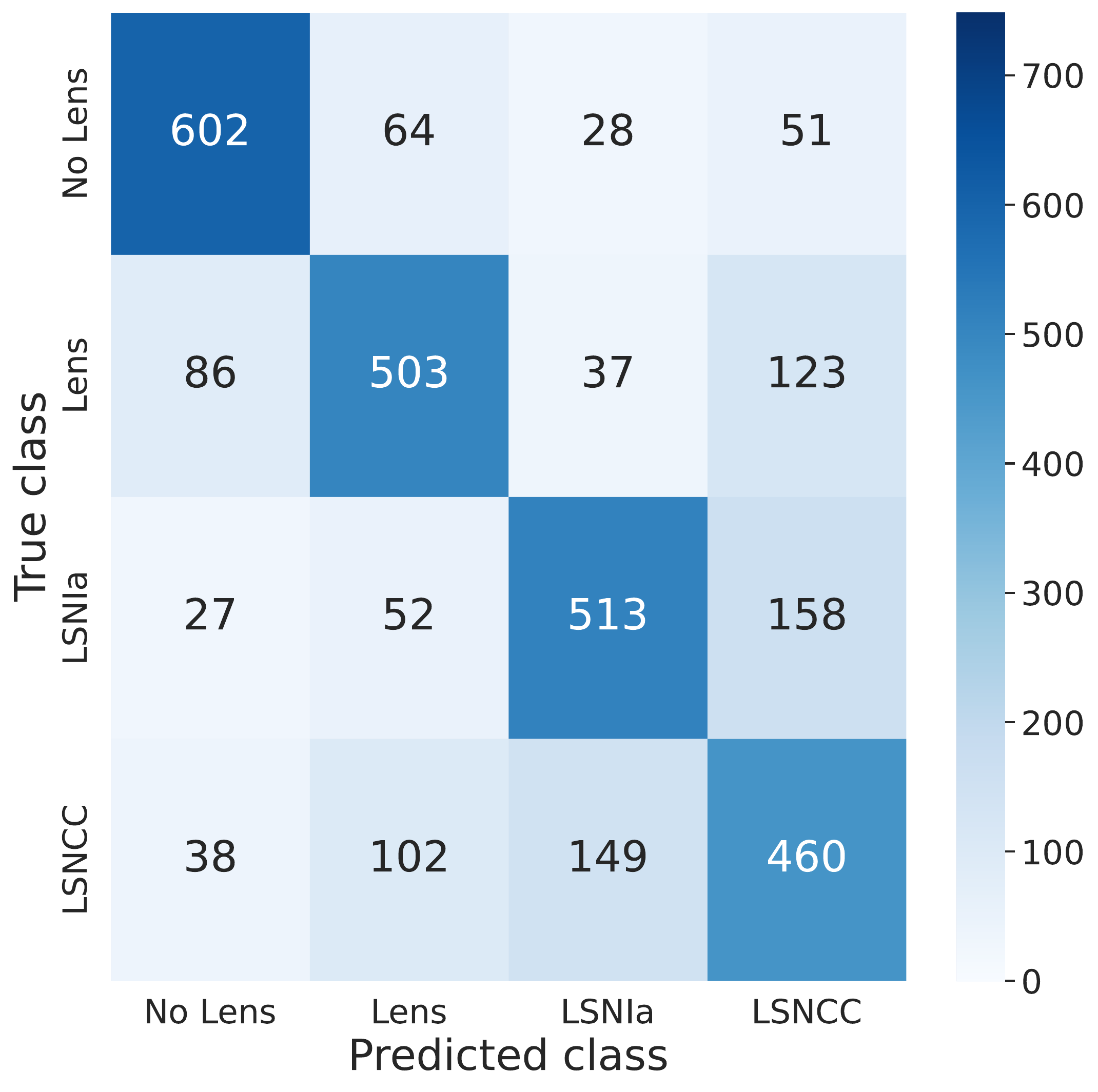}} 
 \subfigure[]{\includegraphics[width=0.45\textwidth]{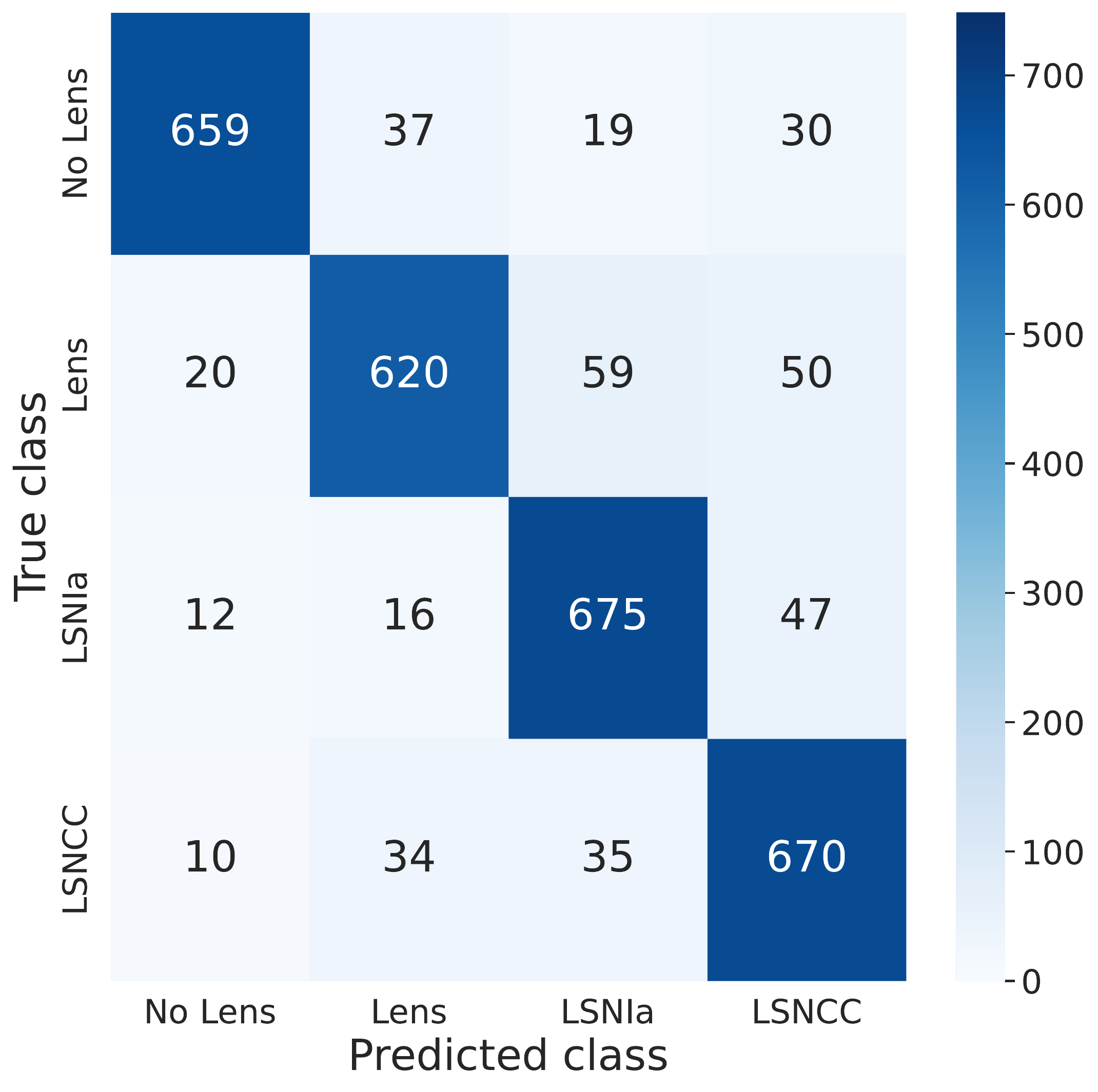}} 
 \subfigure[]{\includegraphics[width=0.45\textwidth]{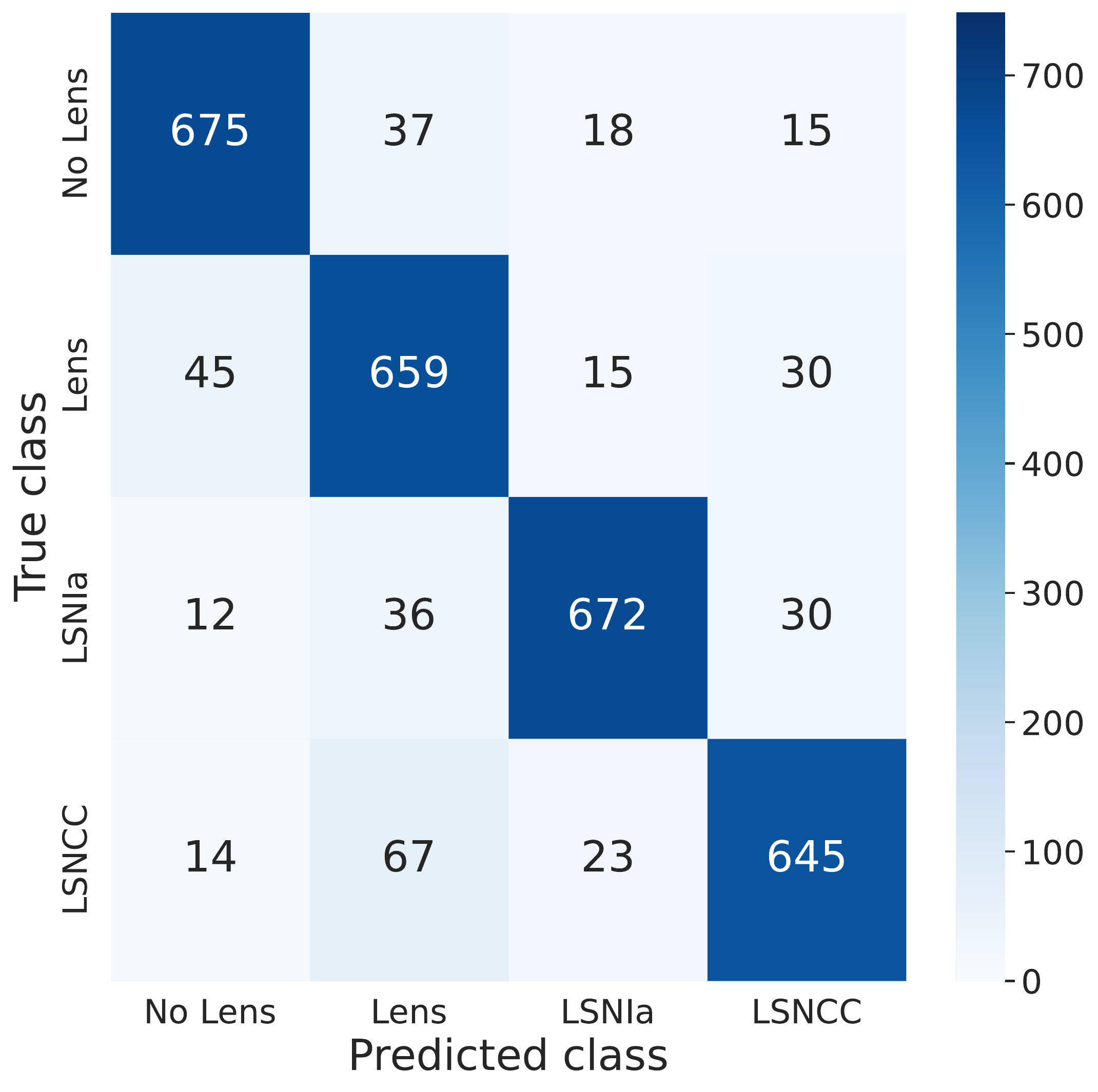}}
 \subfigure[]{\includegraphics[width=0.45\textwidth]{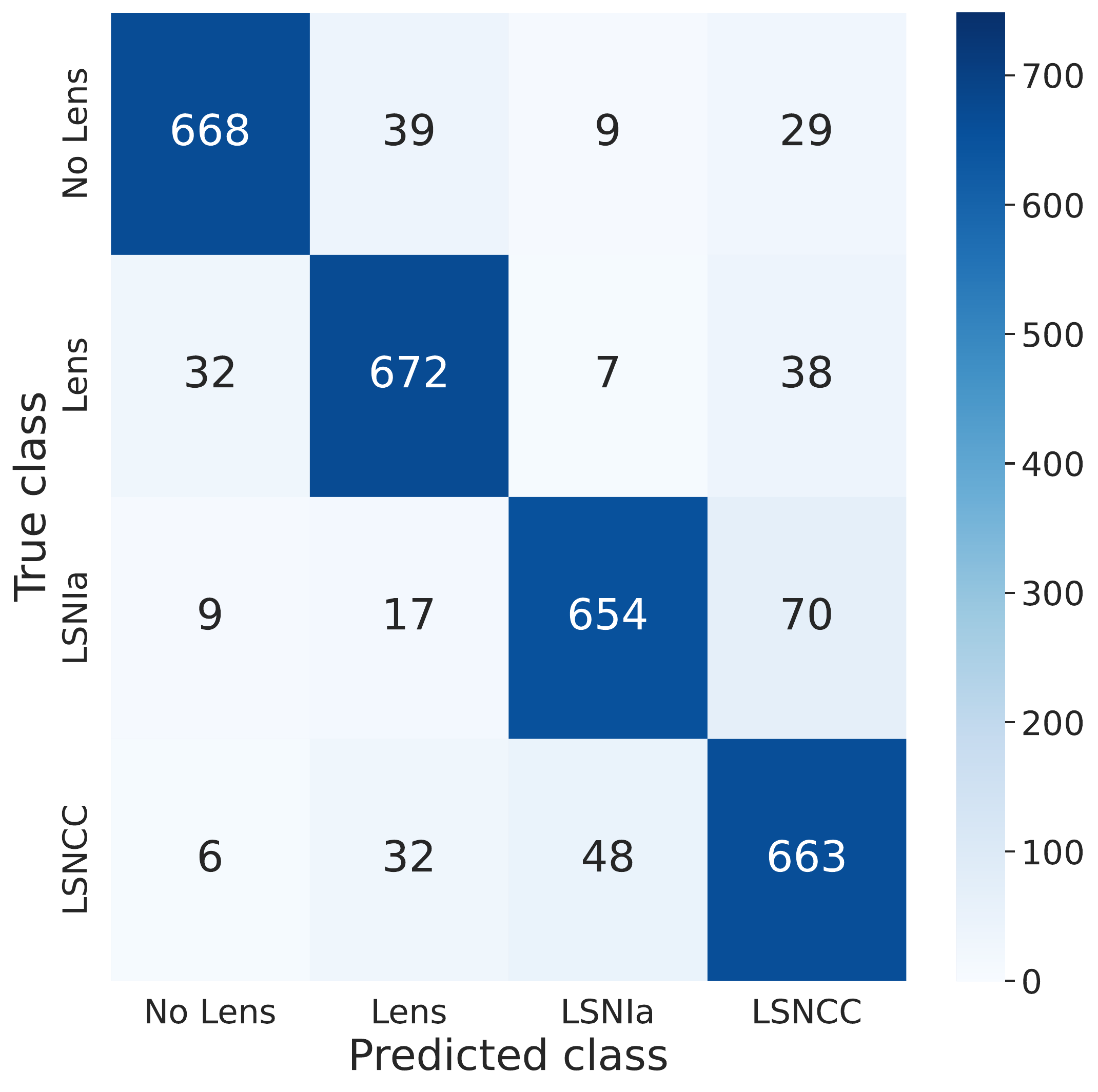}}
 \caption{\textbf{Confusion matrices} of the (a) DES-deep, (b) DES-wide, (c) DESI-DOT, and (d) LSST-wide data sets. In general, the greatest confusion is observed between "Lens" and "No Lens", and in the case of the DES-wide data set, between "LSNCC" and "LSNIa", due to the low sampling rate}
\label{fig:confusion_matrices}
\end{figure*}

Figure \ref{fig:confusion_matrices} illustrates the confusion matrices for the four data sets. For the DES-deep data set, the greatest confusion is observed between "LSNIa" and "LSNCC". A similar, yet more accentuated pattern, was found in \cite{morgan2022deepzipper} too. 

For the DES-wide data set, the confusions between classes are similar, different from \cite{morgan2022deepzipper}, in which the greatest confusion is between "LSNIa" and "LSNCC". This demonstrates that \newnetworkname is more effective at discerning between different gravitationally-lensed transient phenomena, reducing the confusion with respect to the baseline \cite{morgan2022deepzipper} significantly.

For the DESI-DOT data set, the confusion between classes is lower than the one presented in \cite{morgan2022deepzipper}. The greatest confusion is between the "No Lens" and the "Lens" classes, which can be justified by the similarity of the brightness time series of some systems. An example is the "Galaxy + Star" system, in which a galaxy and a star appear close together but without the lensing effect, and the "Galaxy-Galaxy Lensing + Star" system, in which a galaxy stands in front of another galaxy producing the lensing effect and a star appears close to the lensed galaxy from the point of view of the observer.

For the LSST-wide data set the greatest confusion is between the "LSNIa" and the "LSNCC" classes as in DES-deep, similarly to the pattern observed in \cite{morgan2022deepzipper}.

The reported results prove that \newnetworkname can classify the samples of all the data sets accurately and with a significant performance improvement with respect to the compared methods. The results on DES-wide show a significant improvement, reducing the confusion between lensed supernovae classes. This data set is particularly challenging because lensed galaxies are fainter due to the simulated optical depth of the images, which depends on the technical characteristics of the simulated instrumentation. Moreover, the time series are shorter than in the other data sets and thus contain less information.

\subsubsection{Ablation studies}

\begin{table}[!htbp]
\centering
\caption{Comparison of 10 ensemble methods accuracies. The underlined results are the best ones for each data set. The values in bold are the ones comprised in the 1$\sigma$ confidence interval of the best results. The best performances are obtained using SVM on DESI-DOT, DES-deep, and DES-wide, while Max is the best on LSST-wide}
\label{tab:ensemble-comparison}
\resizebox{\textwidth}{!}{%
\begin{tabular}{@{}lccccc@{}}
\toprule
\textbf{Ensemble method} & \textbf{DESI-DOT} & \textbf{DES-deep} & \textbf{DES-wide} & \textbf{LSST-wide} & \textbf{Average} \\ \midrule
\textbf{AdaBoost} & 87.2 $\pm$ 0.6 & 66.7 $\pm$ 0.9 & 86.1 $\pm$ 0.6 & 87.9 $\pm$ 0.6 & 82.0 \\
\textbf{Random Forest} & \textbf{88.3 $\pm$ 0.6} & 68.6 $\pm$ 0.8  & 87.0 $\pm$ 0.6 & \textbf{88.8 $\pm$ 0.6} & 83.2 \\
\textbf{Extra Trees} & \textbf{88.3 $\pm$ 0.6} & 68.7 $\pm$ 0.8 & \textbf{87.1 $\pm$ 0.6} & \textbf{88.8 $\pm$ 0.6} & 83.2 \\
\textbf{Fuzzy ranking} \cite{manna2021fuzzy} & \textbf{88.3 $\pm$ 0.6} & \textbf{68.8 $\pm$ 0.8} & \textbf{87.2 $\pm$ 0.6} & \textbf{88.6 $\pm$ 0.6} & 83.2 \\
\textbf{Average} & \textbf{88.1 $\pm$ 0.6} & \textbf{68.8 $\pm$ 0.8} & \textbf{87.3 $\pm$ 0.6} & \textbf{88.7 $\pm$ 0.6} & 83.2 \\
\textbf{MLP} & 88.0 $\pm$ 0.6 & \textbf{68.9 $\pm$ 0.8} & \textbf{87.6 $\pm$ 0.6} & \textbf{88.5 $\pm$ 0.6} & 83.3 \\
\textbf{KNN} & \textbf{88.3 $\pm$ 0.6} & 68.7 $\pm$ 0.8 & \textbf{87.1 $\pm$ 0.6} & \textbf{88.9 $\pm$ 0.6} & 83.3 \\
\textbf{FCNN} & \textbf{88.4 $\pm$ 0.6} & \textbf{69.2 $\pm$ 0.8} & \textbf{87.6 $\pm$ 0.6} & 88.4 $\pm$ 0.6 & 83.4 \\
\textbf{Max} & \textbf{88.6 $\pm$ 0.6} & 68.7 $\pm$ 0.8 & \textbf{87.3 $\pm$ 0.6} & {\ul \textbf{89.1 $\pm$ 0.6}} & 83.4 \\
\textbf{SVM} & {\ul \textbf{88.7 $\pm$ 0.6}} & {\ul \textbf{69.6 $\pm$ 0.8}} & {\ul \textbf{87.7 $\pm$ 0.6}} & \textbf{88.8 $\pm$ 0.6} & 83.7 \\ \midrule
\textbf{Improvement w.r.t. \munetnospace} & 0.8 & 1.7 & 1.2 & 0.3 & 1.0 \\ \bottomrule
\end{tabular}%
}
\end{table}

\begin{table}[!htbp]
\centering
\caption{\textbf{Ablation studies on SVM ensemble} -- When a single network is considered, accuracy refers to the results obtained by applying it without any additional decision-level algorithm. The underlined results are the best mean accuracy results for every data set, and results in bold are contained within the confidence intervals of the best results. All the values are expressed in \%}
\label{tab:ablation}
\resizebox{0.8\textwidth}{!}{%
\begin{tabular}{@{}ccccc@{}}
\toprule
\textbf{Data set} & \textbf{\lonetnospace} & \textbf{\glonetnospace} & \textbf{\munetnospace} & \textbf{Accuracy $\pm$ 1$\sigma$} \\ \midrule
\multirow{7}{*}{DESI-DOT} & \checkmark &  &  & 87.0 $\pm$ 0.6 \\
 &  & \checkmark &  & 77.2 $\pm$ 0.8 \\
 &  &  & \checkmark & 87.9 $\pm$ 0.6 \\
 & \checkmark & \checkmark &  & 87.0 $\pm$ 0.6 \\
 & \checkmark &  & \checkmark & {\ul \textbf{88.7 $\pm$ 0.6}} \\
 &  & \checkmark & \checkmark & 87.9 $\pm$ 0.6 \\
 & \checkmark & \checkmark & \checkmark & {\ul \textbf{88.7 $\pm$ 0.6}} \\ \midrule
\multirow{7}{*}{DES-deep} & \checkmark &  &  & 67.5 $\pm$ 0.9 \\
 &  & \checkmark &  & 62.3 $\pm$ 0.9 \\
 &  &  & \checkmark & 67.9 $\pm$ 0.9 \\
 & \checkmark & \checkmark &  & 68.4 $\pm$ 0.8 \\
 & \checkmark &  & \checkmark & 68.7 $\pm$ 0.8 \\
 &  & \checkmark & \checkmark & 68.7 $\pm$ 0.8 \\
 & \checkmark & \checkmark & \checkmark & {\ul \textbf{69.6 $\pm$ 0.8}} \\ \midrule
\multirow{7}{*}{DES-wide} & \checkmark &  &  & 85.8 $\pm$ 0.6 \\
 &  & \checkmark &  & 76.8 $\pm$ 0.8 \\
 &  &  & \checkmark & 86.5 $\pm$ 0.6 \\
 & \checkmark & \checkmark &  & \textbf{87.2 $\pm$ 0.6} \\
 & \checkmark &  & \checkmark & \textbf{87.3 $\pm$ 0.6} \\
 &  & \checkmark & \checkmark & 86.9 $\pm$ 0.6 \\
 & \checkmark & \checkmark & \checkmark & {\ul \textbf{87.7 $\pm$ 0.6}} \\ \midrule
\multirow{7}{*}{LSST-wide} & \checkmark &  &  & 87.2 $\pm$ 0.6 \\
 &  & \checkmark &  & 76.8 $\pm$ 0.8 \\
 &  &  & \checkmark & \textbf{88.5 $\pm$ 0.6} \\
 & \checkmark & \checkmark &  & 87.4 $\pm$ 0.6 \\
 & \checkmark &  & \checkmark & \textbf{88.5 $\pm$ 0.6} \\
 &  & \checkmark & \checkmark & \textbf{88.5 $\pm$ 0.6} \\
 & \checkmark & \checkmark & \checkmark & {\ul \textbf{88.8 $\pm$ 0.6}} \\ \bottomrule
\end{tabular}%
}
\end{table}

Table \ref{tab:ensemble-comparison} compares SVM with  other ensemble methods. The use of SVM brings an average 1\% improvement over the best multi-modal network (\munetnospace) and  surpasses the performances of other ensemble methods in three data sets out of four. Considering the LSST-wide data set, Max performs better than SVM, but the SVM result is inside Max's confidence interval. Moreover, Max's accuracy on DES-deep  is outside the SVM confidence interval. Considering the analyzed ensemble methods, only SVM, Fuzzy Ranking \cite{manna2021fuzzy} and Average are inside the  confidence interval of the best ensemble approach for all the data sets. However, both Fuzzy Ranking and Average have an accuracy significantly inferior to that of SVM.

Table \ref{tab:ablation} presents the results of the ablation experiments with respect to the multi-modal sub-networks. The presence of the three sub-networks  guarantees the highest accuracy, with the results obtained ensembling one or two networks being often outside the confidence interval of the result obtained by ensembling three networks. In particular, combining three networks yields an improvement ranging from $+0.3\%$ to $+12.0\%$ with respect to single networks, and a change ranging from $0.0\%$ to $+1.7\%$ with respect to the combination of two networks. 

In DESI-DOT, the contribution of \glonet is dominated by that of the other two sub-networks and thus eliminating \glonet does not affect accuracy. This can be explained by the use of early fusion in \glonet which does not preserve the information of the image, which is immediately fused with the time series.

The introduction of the means $\mu$ and standard deviations $\sigma$ of the time series yields an additional modest average improvement of $0.5\%$ in accuracy consistently across the data sets. Compared to the predictions made using a random forest with inputs $\mu$ and $\sigma$, \newnetworkname accuracy improves from $18\%$ to $49\%$. 

\subsubsection{Execution time}

\newnetworkname has been trained using an NVIDIA GeForce GTX 1080 Ti for \glonetnospace, \munet and \lonetnospace. On average, the network training requires less than 3 hours for a single data set. SVM training time is negligible with respect to the other networks.

\subsection{Qualitative results}
\label{sec:qualitative-results}

Section \ref{sec:qualitative-simulated} illustrates some representative examples of the results obtained by \newnetworkname on the four test sets. All the images are obtained by adding the \textit{griz} layers, as done in \cite{morgan2022deepzipper}. In the plots, the g band is displayed in green, the r band in red, the i band in blue and the z band in grey.

Section \ref{sec:qualitative-real} shows how the application of  \newnetworkname to real data recognizes the presence of gravitational lensing phenomena, also confirming the three lensed supernovae candidate systems, a very rare occurrence,  reported in \cite{morgan2022deepzipper2}.

\subsubsection{Simulated data}
\label{sec:qualitative-simulated}

\begin{figure*}
 \centering
 \includegraphics[width=0.8\textwidth]{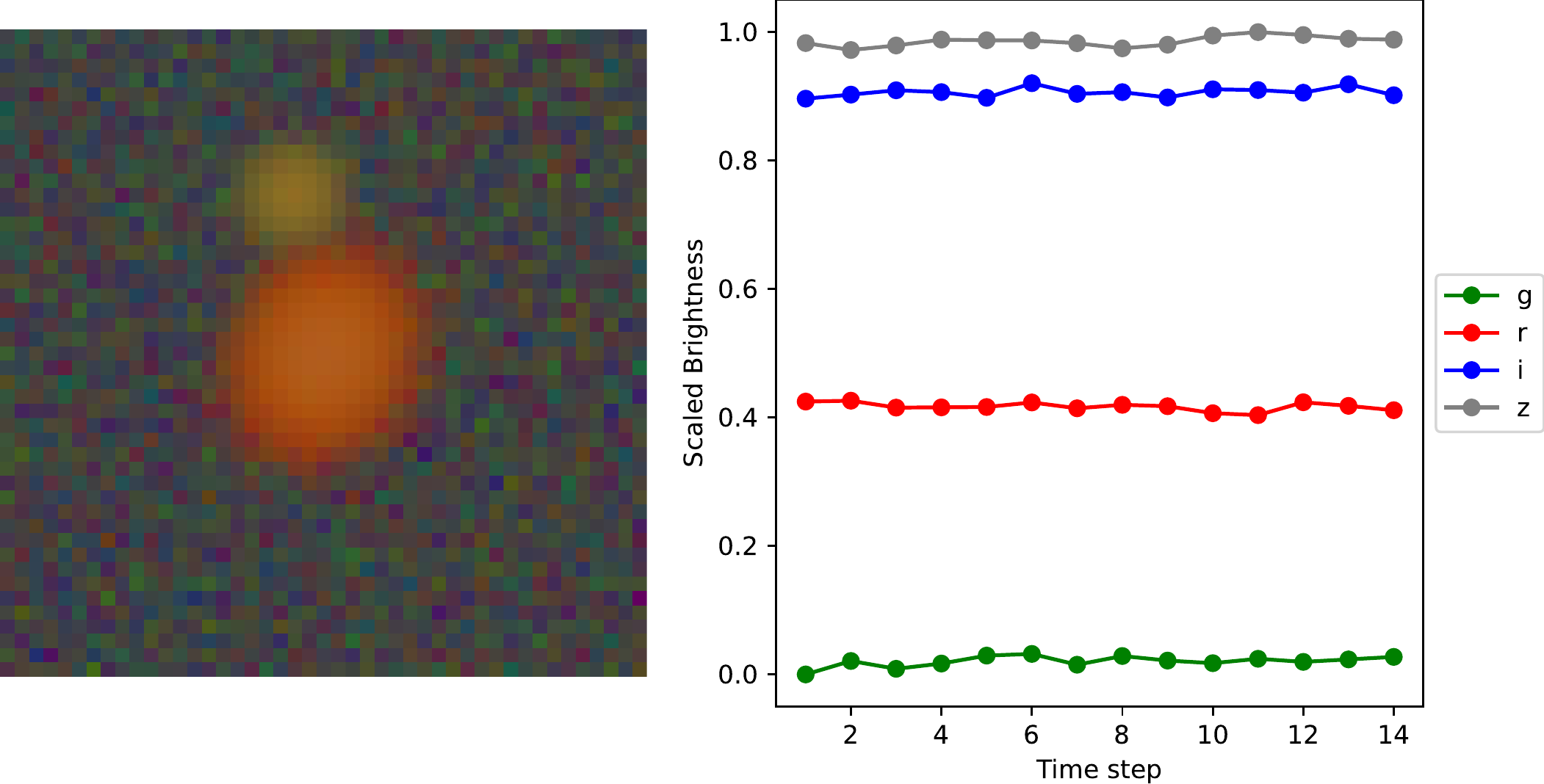}
 \caption{\textbf{A positive example on the LSST-wide data set} -- This datum belongs to the "No Lens" class. The image shows two separate stars that have a spherical geometry, which suggests they are not lensed. Moreover, the curves on the right show no consistent brightness variation through time, which indicates the absence of transient phenomena}
 \label{fig:example-60}
\end{figure*}

Figure \ref{fig:example-60} presents a true positive example belonging to the "No Lens" class in the LSST-wide data set. It shows two stars close to each other, which exhibit a spherical symmetry, which suggests the absence of lensing. In addition, the brightness curves do not show consistent variations, which indicates the absence of transient phenomena.

\begin{figure*}
 \centering
 \includegraphics[width=0.8\textwidth]{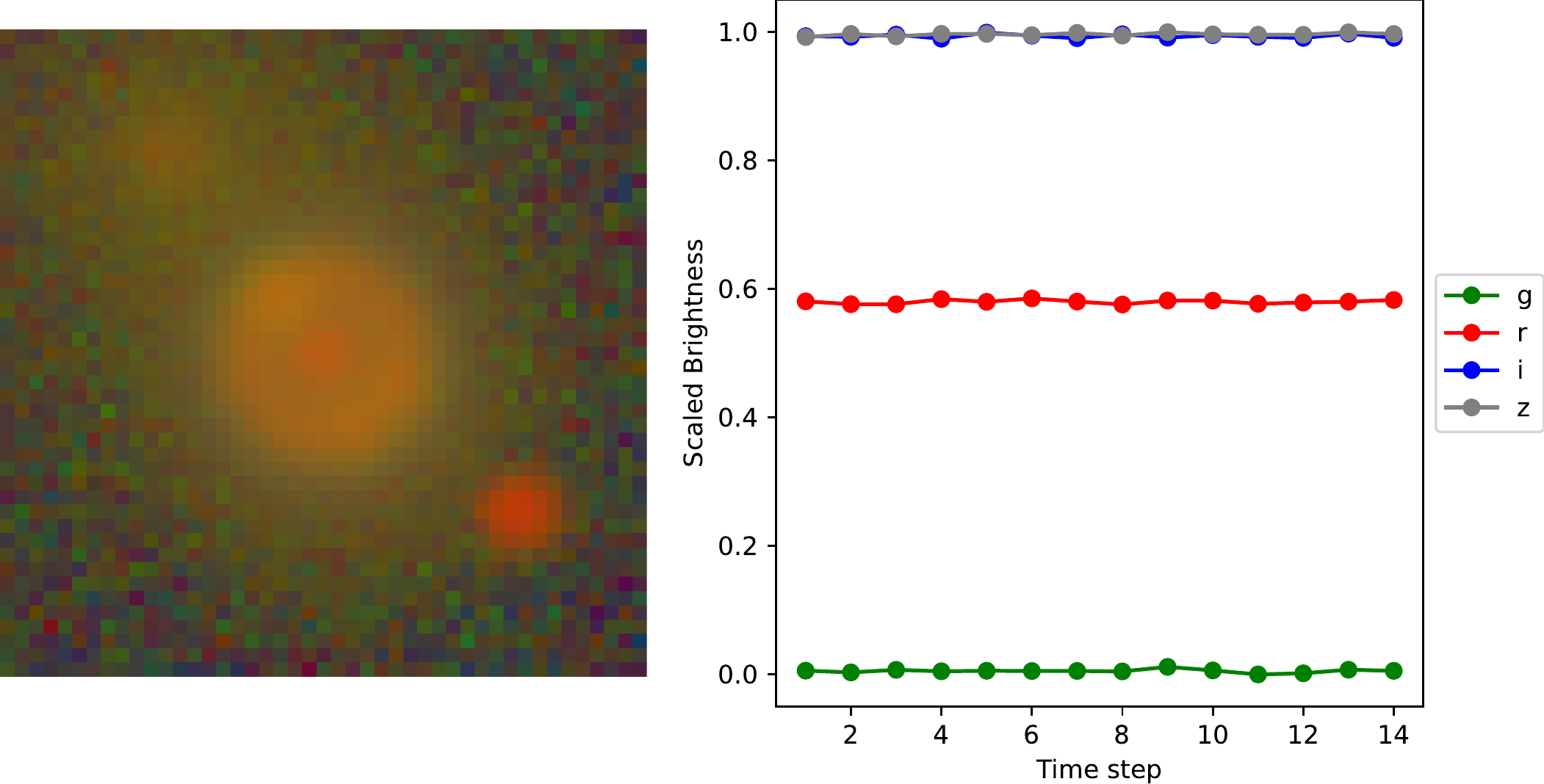}
 \caption{\textbf{A positive example on the DESI-DOT data set} -- This datum belongs to the "Lens" class. The lensing effect is visible in the ring pattern around the central body. The flatness of the brightness time series, instead, indicates the absence of transient phenomena (e.g., explosions), which is expected because the involved entities are galaxies}
 \label{fig:example-174}
\end{figure*}

Figure \ref{fig:example-174} presents a true positive example belonging to the "Lens" class, in the DESI-DOT data set. In this system, the lensing effect is manifested by the ring pattern on the central body. The flatness of the brightness curves indicates the absence of transient phenomena, as expected, because the system is formed by galaxies, which are not characterized by explosive events.

\begin{figure*}
 \centering
 \includegraphics[width=0.8\textwidth]{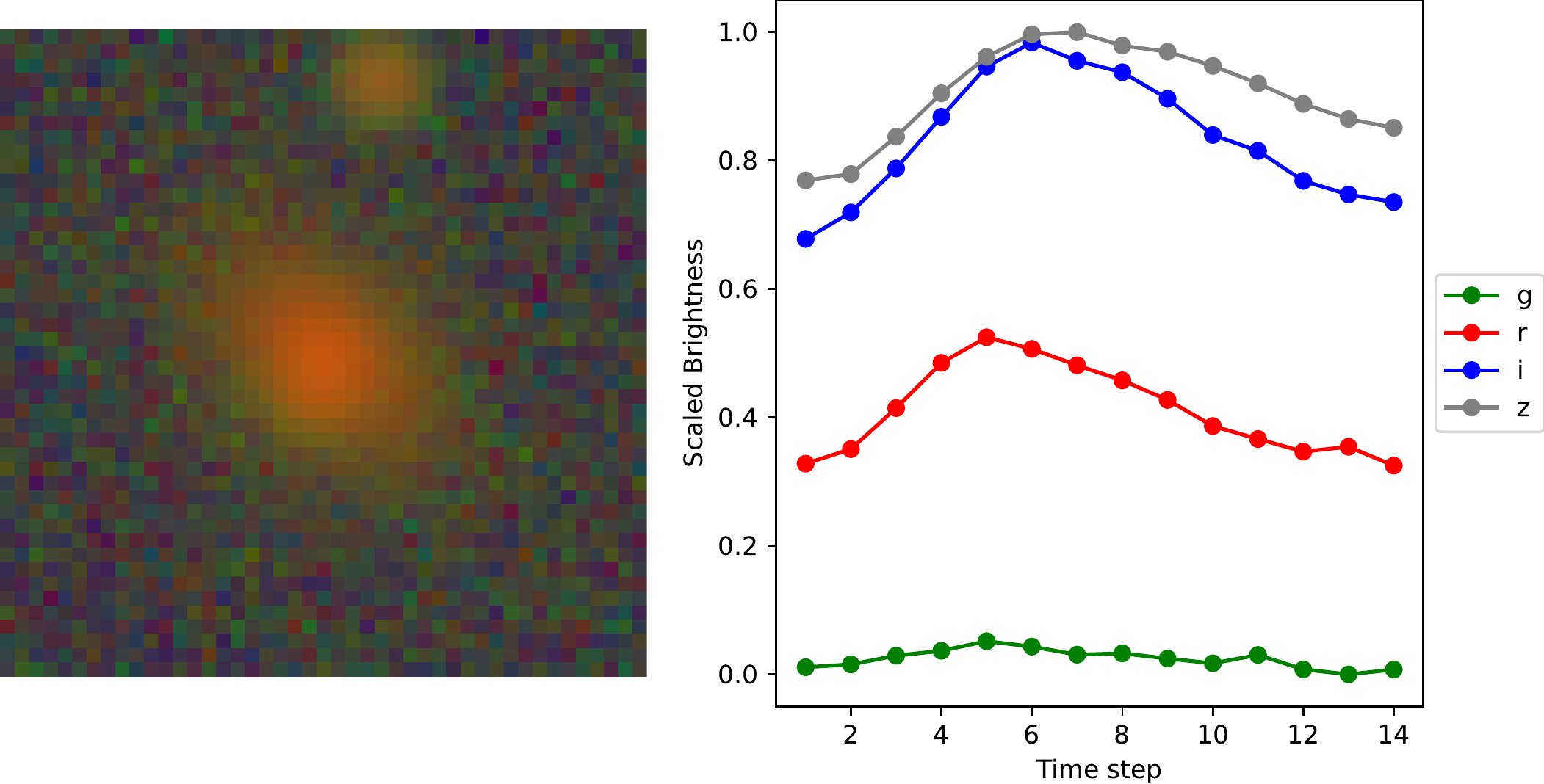}
 \caption{\textbf{A positive example on the DESI-DOT data set} -- This datum belongs to the "LSNIa" class. The lensing effect is visible from the elliptical shape of the central body, while the presence of a supernova can be observed by the peaks in the brightness time series, which indicates the presence of explosive transient phenomena. The supernova type can be inferred from the flatness of the g band time series}
 \label{fig:example-142}
\end{figure*}

Figure \ref{fig:example-142} presents a true positive example belonging to the "LSNIa" class in the DESI-DOT data set. The peak in the time series indicates the presence of an exploding supernova and the image shows an elliptical shape, which signals the presence of lensing. The brightness in the g band is almost flat, which is distinctive of Type Ia supernovae. Type Ia and core-collapse supernovae release chemical elements during the explosion and produce photons at different wavelengths, which are detected by sensors in specific bands. During explosions, the emission of an element with a certain wavelength produces a temporary brightness peak in the corresponding band. Both types of supernovae release chemical elements whose detection can be observed in the g band, but Type-Ia supernovae emit less materials than core-collapse supernovae, which makes the latter exhibit a more pronounced peak in the g band. The absence of such a peak in Figure \ref{fig:example-142} justifies the "LSNIa" classification.

\begin{figure*}
 \centering
 \includegraphics[width=0.8\textwidth]{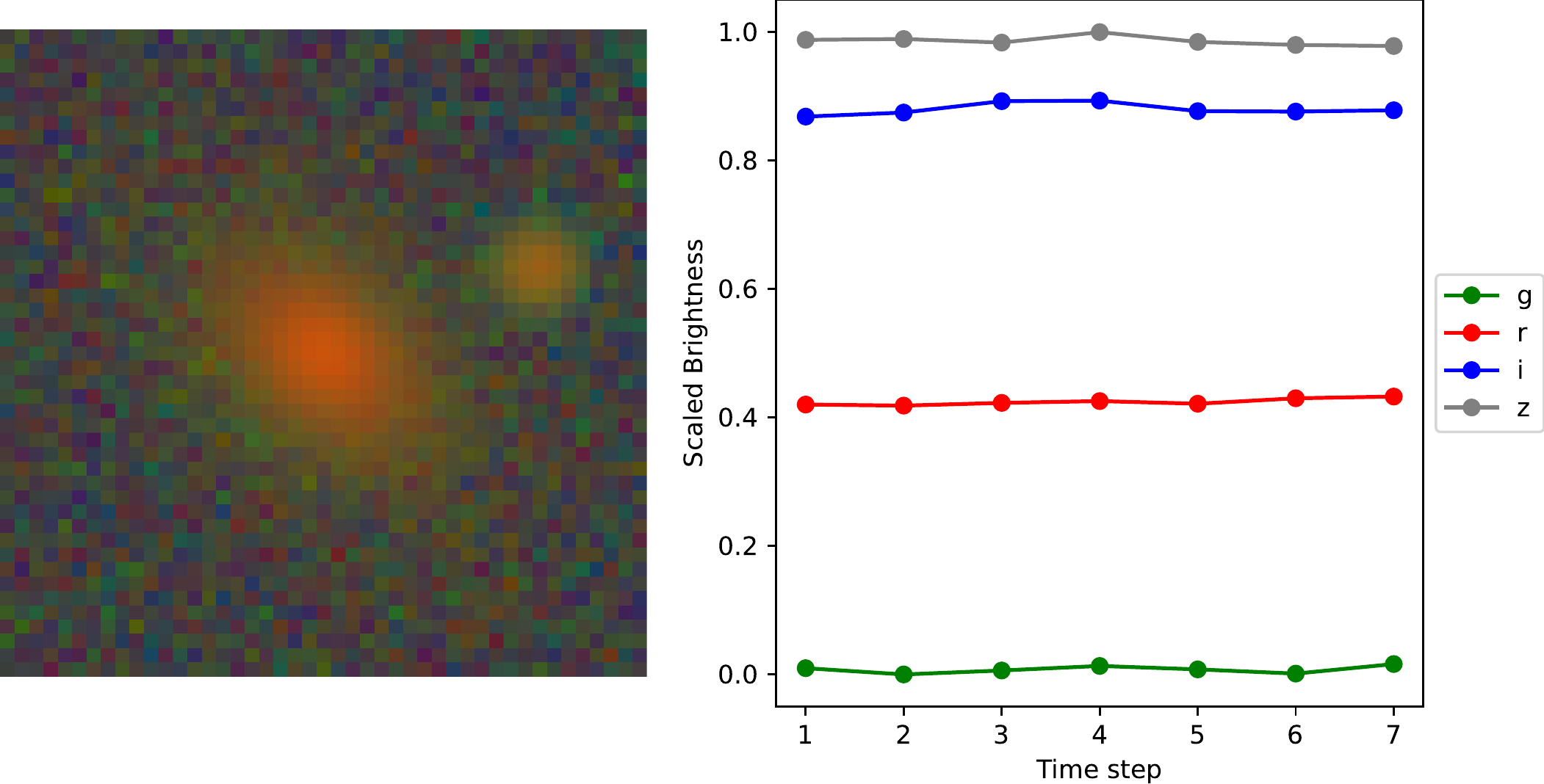}
 \caption{\textbf{A positive example on the DES-wide data set} -- This datum belongs to the "LSNIa" class. The lensing effect is visible because of the elliptical shape of the central body. Even if the peaks that indicate the presence of transient phenomena are absent, the network is still able to correctly classify the datum}
 \label{fig:example-142-full}
\end{figure*}

The same type of system is shown in Figure \ref{fig:example-142-full}, from the DES-wide data set. In this case, the peaks are not detected because of the lower sampling rate, which misses rapid transient events. However, the network correctly classifies this example thanks to the information contained in the image. 

\begin{figure*}
 \centering
 \includegraphics[width=0.8\textwidth]{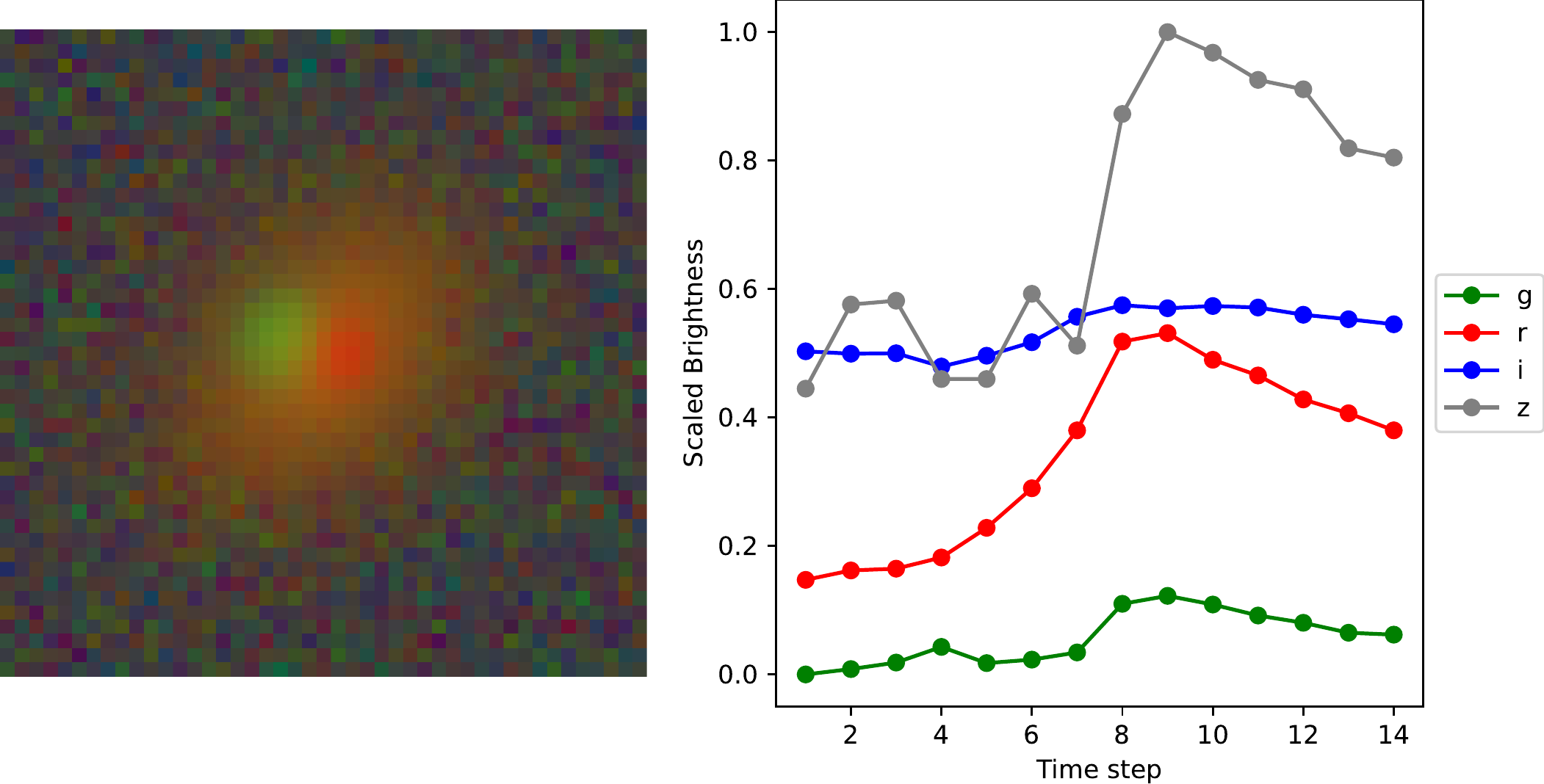}
 \caption{\textbf{A positive example on the DESI-DOT data set} -- This datum belongs to the "LSNCC" class. In this case, the lensing effect is suggested both by the presence of varying time curves (indicating the presence of a supernova) and the green body lensed by the galaxy}
 \label{fig:example-258}
\end{figure*}

Figure \ref{fig:example-258} presents a true positive example belonging to the "LSNCC" class, in the DESI-DOT data set. In this case, the presence of a supernova is indicated by the rapid variation in the brightness time series. Since also the g band exhibits a peak, the input is classified as a core-collapse supernova. The lensing effect is manifested in the image by the supernova (the green body), lensed by the galaxy in front of it. The green color confirms the presence of elements emitting photons in the g band and the body itself is visible because of the magnifying effect induced by the galaxy.

\begin{figure*}
 \centering
 \includegraphics[width=0.8\textwidth]{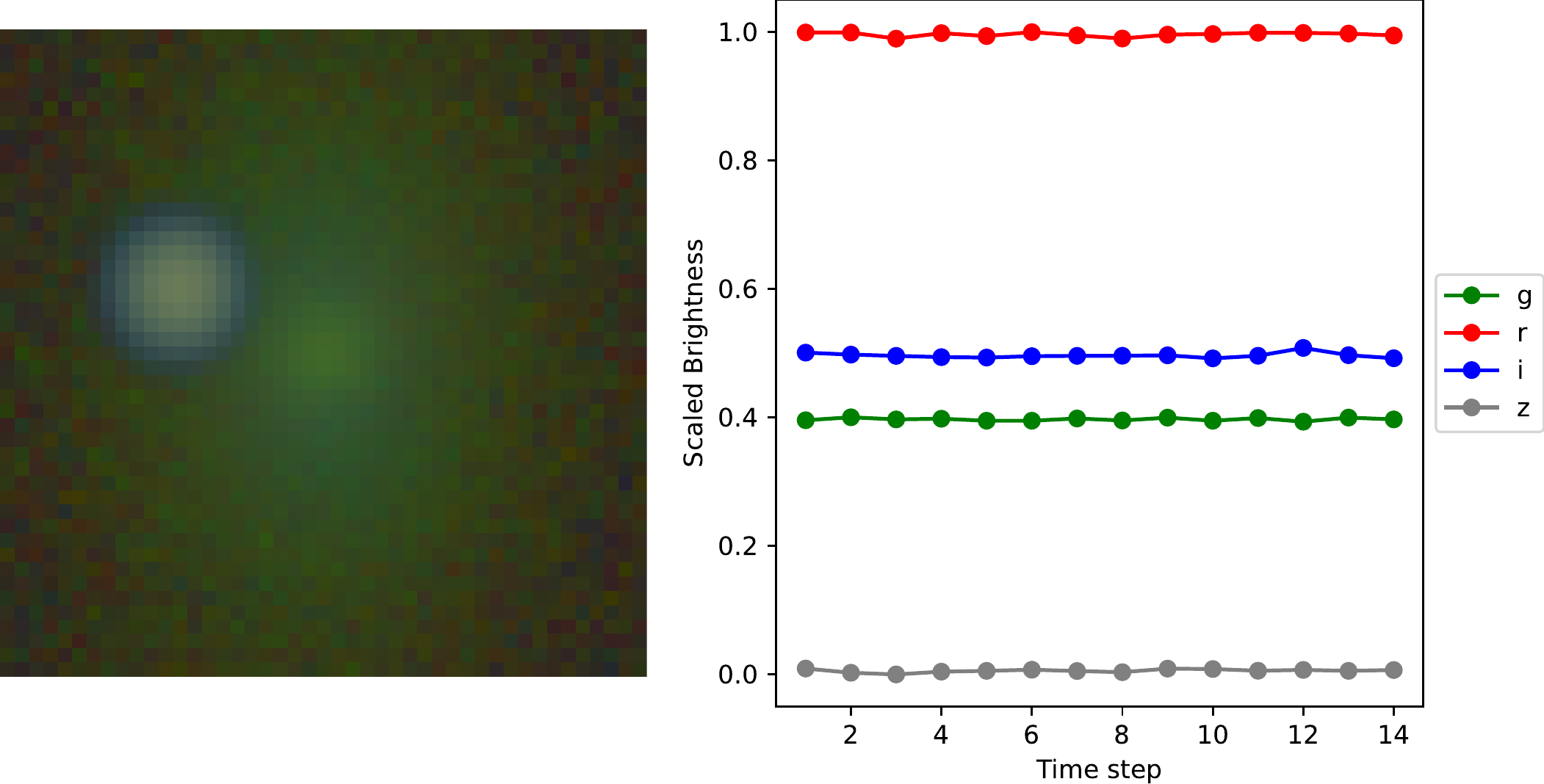}
 \caption{\textbf{A negative example on the LSST-wide data set} -- This datum belongs to the "LSNCC" class, but has been classified as "Lens". The lensing effect is alluded by the halo surrounding the star, while the flat time series suggests the absence of a transient phenomenon, which induces the wrong classification}
 \label{fig:example-143}
\end{figure*}

Figure \ref{fig:example-143} presents a negative example in the LSST-wide data set. The datum belongs to the "LSNCC" class, but is classified as "Lens", which means that the model was not able to detect the presence of a supernova and interpreted the example as a lensed system without evident transient phenomena. The wrong classification is caused by the low-quality time series and the ambiguous image. The lensing effect is visible thanks to the faint halo surrounding the star in the background, but the time series (wrongly) suggest the absence of a transient phenomenon. The apparent lack of the transient phenomenon can be explained by considering that supernovae explosions can happen in a short time and the brightness variation may not be recorded by the camera. Soon after the explosion, the brightness returns to the original value, which explains the flatness of the curves.

\begin{figure*}
 \centering
 \includegraphics[width=0.8\textwidth]{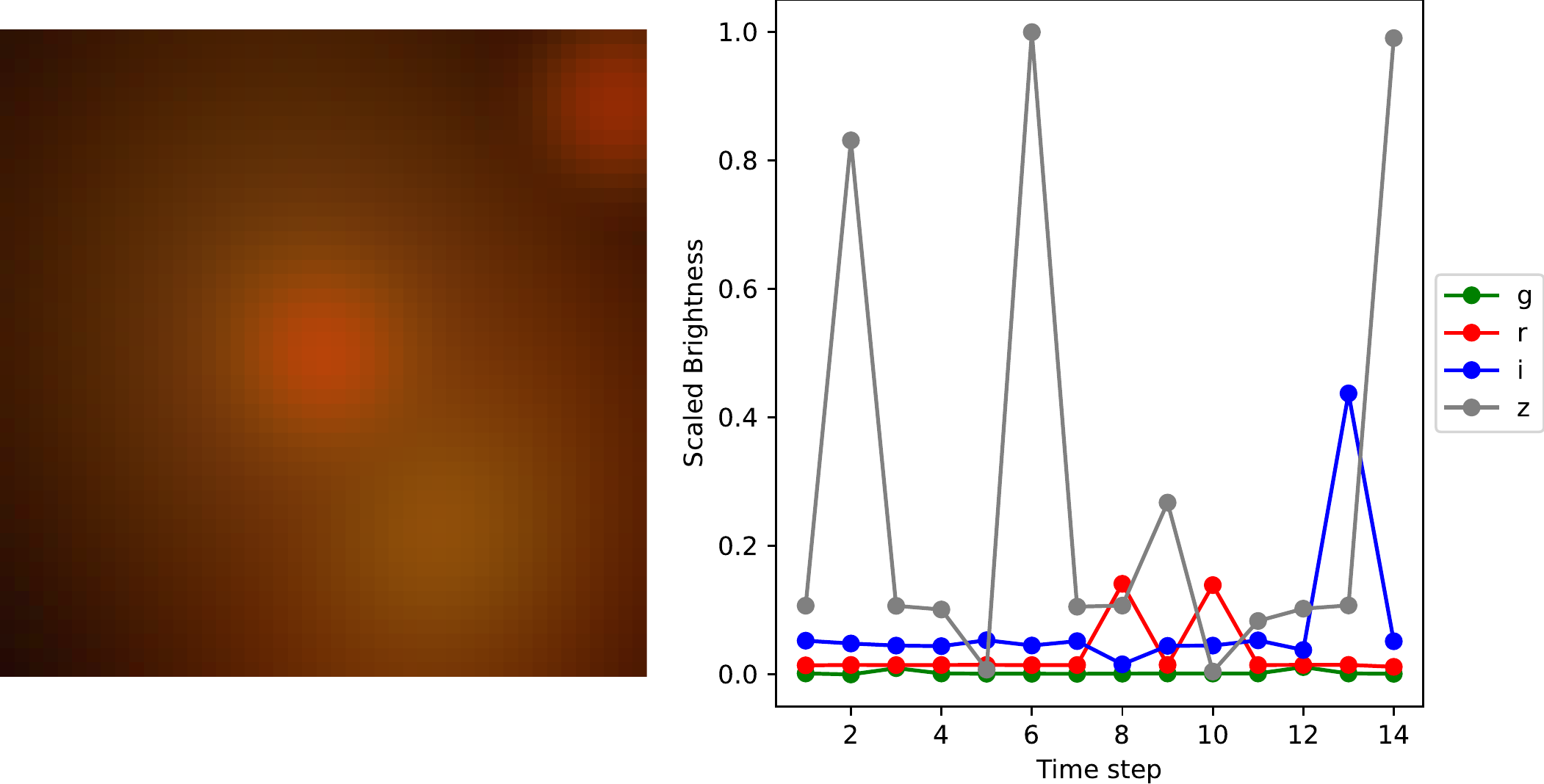}
 \caption{\textbf{A negative example on the DES-deep data set} -- This datum belongs to the "Lens" class, but has been classified as "No Lens". The lensing effect is suggested by the halo surrounding the central body}
 \label{fig:example-2437}
\end{figure*}

Figure \ref{fig:example-2437} presents a negative example from the DES-deep data set, belonging to the "Lens" class, but classified as "No Lens".  The lensing effect is visible on the central body, which has a halo. However, because of the low image resolution, this effect is not as clear in most of the positive examples. In addition, the presence of multiple peaks is not frequently associated with the "Lens" class and induces the wrong classification.

\begin{figure*}
 \centering
 \includegraphics[width=0.8\textwidth]{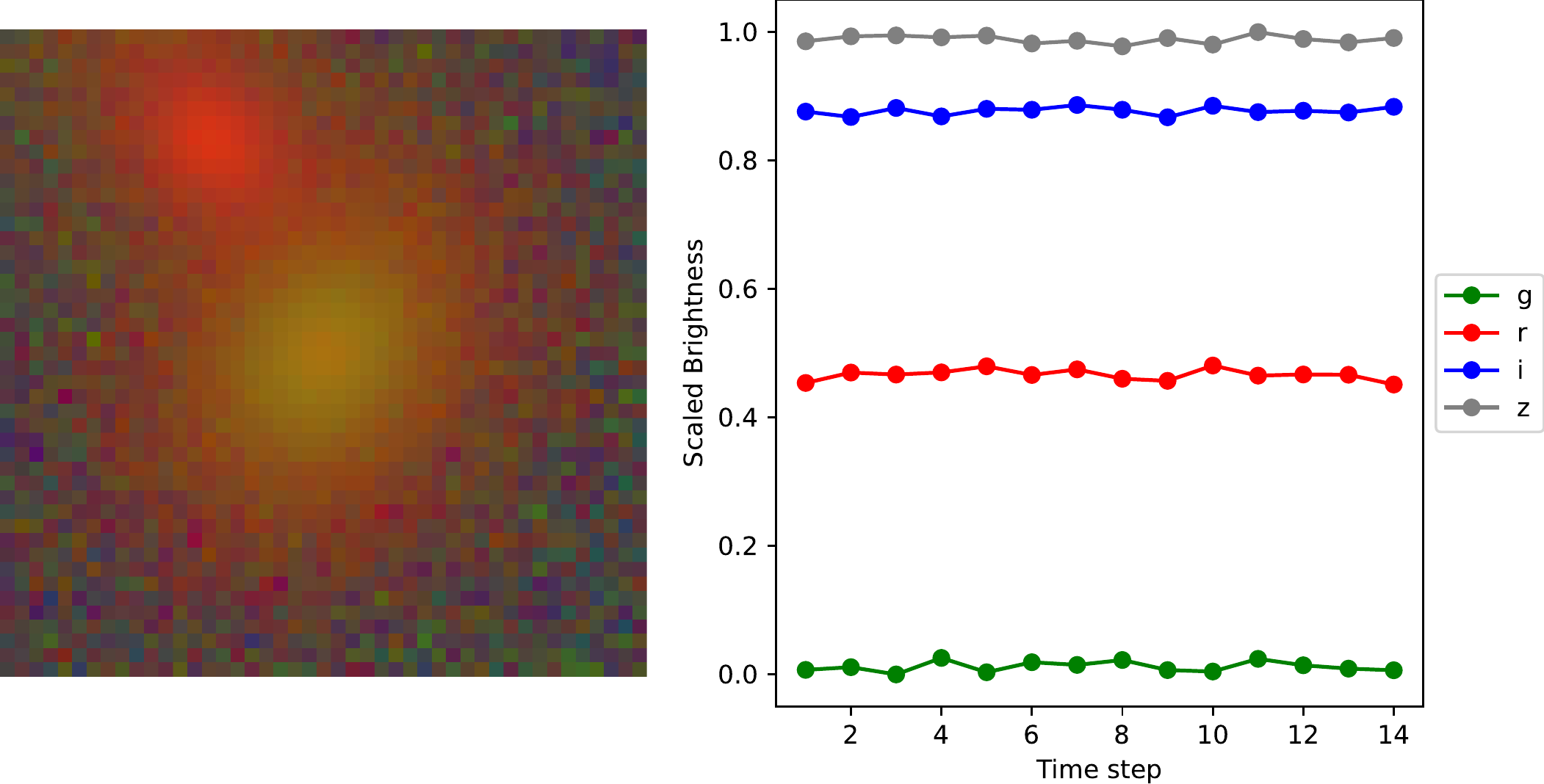}
 \caption{\textbf{A negative example on the DESI-DOT data set} -- This datum belongs to the "No Lens" class, but it has been classified as belonging to the "Lens" class. The lensing effect is suggested by the elliptical shape, but such shape may suggest also the presence of a non-lensing elliptical galaxy. The flatness of the time series, in addition, does not allow to discern "Lens" and "No Lens" systems, as some "No Lens" systems also have flat time series}
 \label{fig:example-905}
\end{figure*}

As a final example, Figure \ref{fig:example-905} shows an ambiguous image in the DESI-DOT data set, incorrectly classified. The sample belongs to the "No Lens" class, but is classified as "Lens". The confusion is generated chiefly by the elliptical  object, which is confused with a lensing effect, while it can represent, e.g., a non-lensed elliptical galaxy. The time series are flat, so they do not help discern "Lens" and "No Lens" systems, because some "No Lens" systems also have flat time series.

\subsubsection{Real data}
\label{sec:qualitative-real}

The authors of \cite{morgan2022deepzipper2} analyze real data from the Dark Energy Survey over a five-year period (Y1-Y5) with the aim of detecting gravitationally-lensed supernovae. They identify three potential lensed supernova systems (identified as 691022126, 701263907, and 699919273), two of which were detected using only Y5 data, indicating that the supernovae likely exploded during that year. Our research tries to reproduce such results using public data provided by NoirLab\footnote{\url{https://datalab.noirlab.edu/} (As of March 2023)}, which currently only includes data up to Y4, using the network trained on the DES-deep data set.

\newnetworkname  successfully identified the lensed supernova with ID 691022126 and also detected the presence of a gravitational lens for the other two systems. To extract brightness time series, we followed a  methodology similar to the one employed in \cite{morgan2022deepzipper2}, using 14 time steps with a 6-day interval between each step, resulting in a 78-day period. The corresponding image was obtained by averaging the images captured during this period. Each system has been observed for more than 78 days, and as such, multiple observations are associated with each system. Finally, images bigger than $45 \times 45$ pixels are resized to such dimension.

Table \ref{tab:real-systems} presents a summary of our results on the real data. The number of observations associated with each system may differ slightly due to missing observations in the database. Our results confirm the findings of \cite{morgan2022deepzipper2}. The systems in which a lensed supernova was discovered only in Y5 have a prevalence of "Lens" prediction.

\begin{table}[!htbp]
\centering
\caption{Summary of results on the considered real data, including system ID, coordinates, number of observations, predicted class, and the proportion of observations in which that class was observed. Here, RA indicates the right ascension, and DEC indicates the declination}
\label{tab:real-systems}
\resizebox{\textwidth}{!}{%
\begin{tabular}{@{}cccccc@{}}
\toprule
\multirow{2}{*}{\textbf{System ID} \cite{morgan2022deepzipper2}} & \multicolumn{2}{c}{\textbf{Coordinates [deg]}} & \multirow{2}{*}{\textbf{\# observations}} & \multicolumn{2}{c}{\textbf{Predicted class}} \\ \cmidrule(lr){2-3} \cmidrule(l){5-6} 
 & \textbf{RA} & \textbf{DEC} &  & \textbf{Class} & \textbf{Proportion} \\ \midrule
691022126 & 53.898910 & -28.912293 & 77 & LSNCC & $65\%$ \\
701263907 & 40.969218 & -0.619054 & 69 & Lens & $100\%$ \\
699919273 & 10.155917 & -44.437515 & 76 & Lens & $93\%$ \\ \bottomrule
\end{tabular}%
}
\end{table}

\begin{figure*}
 \centering
 \includegraphics[width=0.8\textwidth]{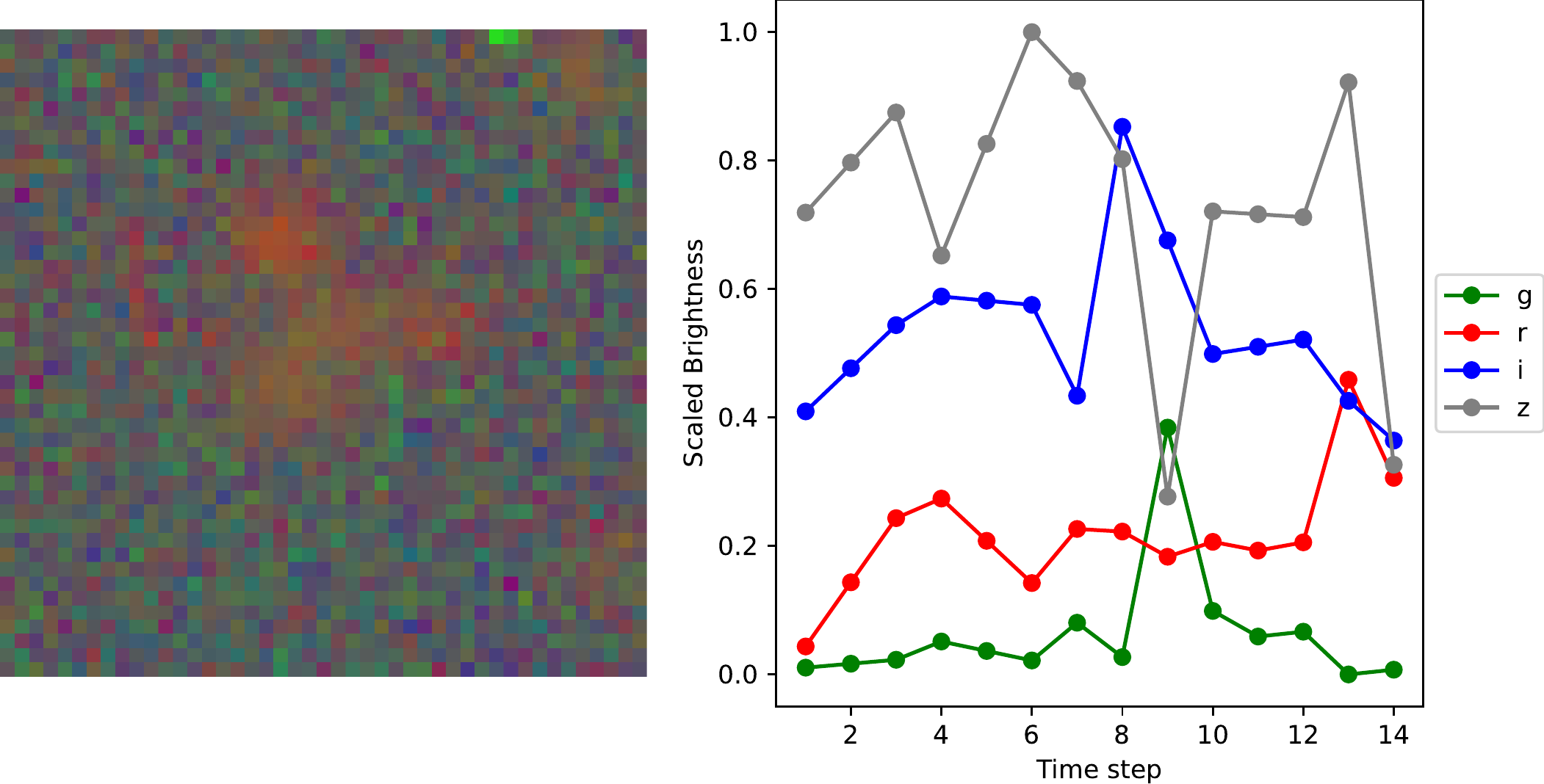}
 \caption{\textbf{The detection of a real gravitationally-lensed supernova} -- This system is formed by four objects, whose boundaries are not well-defined. The time series shows the presence of peaks in the four bands. The presence of a peak in the g band suggests the presence of a LSNCC, as predicted by \newnetworknamenospace}
 \label{fig:example-691022126-supernova}
\end{figure*}

\begin{figure*}
 \centering
 \includegraphics[width=0.8\textwidth]{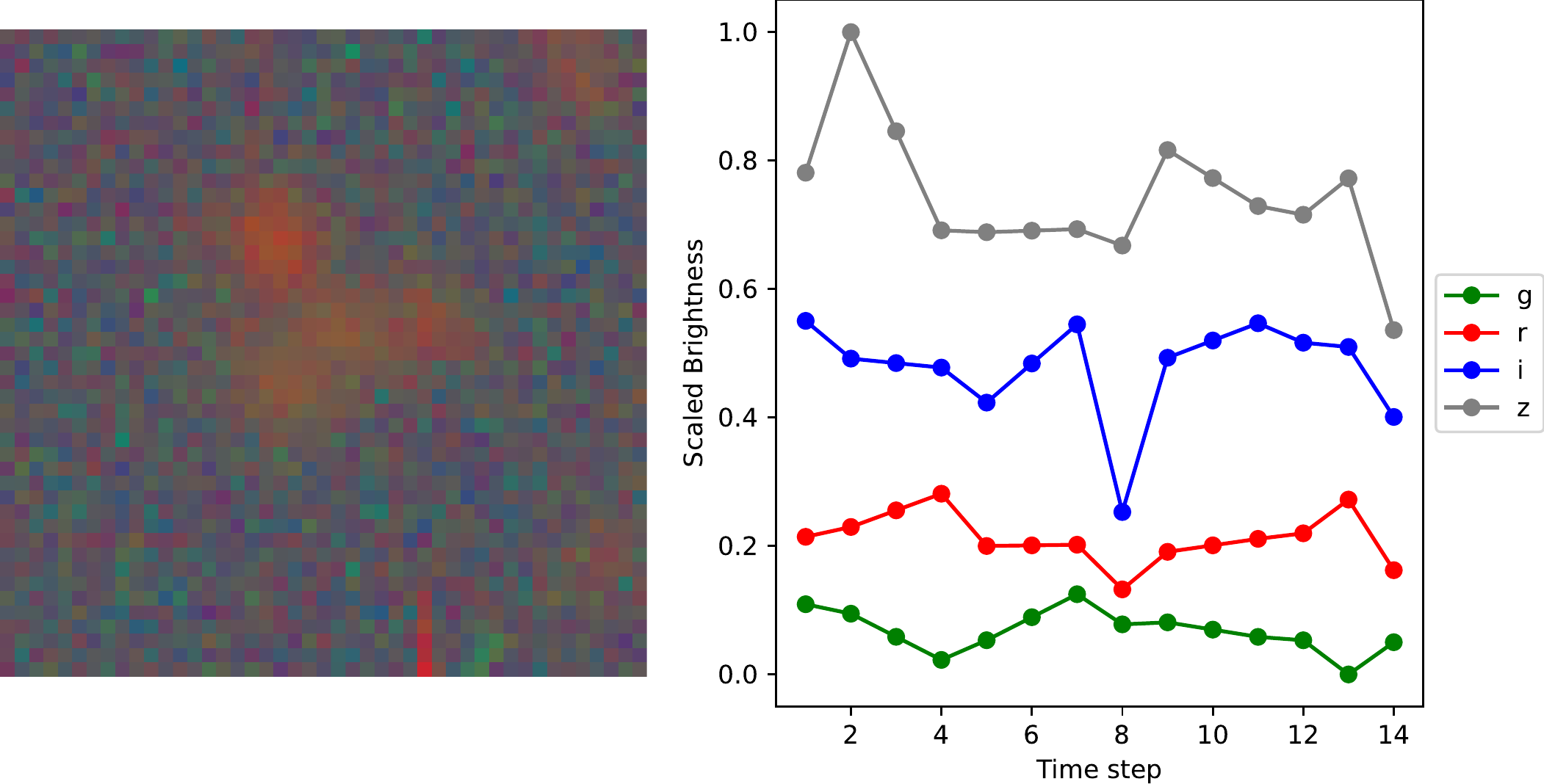}
 \caption{\textbf{The missed detection of a real gravitationally-lensed supernova} -- The system presented in this figure is the same as the one in Figure \ref{fig:example-691022126-supernova}, but the time series, for this time interval, does not show significant peaks, suggesting the absence of a transient phenomenon. The clearer separation between the four bodies in the image is not enough for suggesting the presence of a supernova}
 \label{fig:example-691022126-flat}
\end{figure*}

The object with ID 691022126 is shown in Figure \ref{fig:example-691022126-supernova}. It has been classified as "LSNCC" in $65\%$ of the observations. The presence of a gravitational lens    is signaled by the multiple objects visible in the image. Additionally, the peaks in the four bands indicate the presence of a supernova and the peak in the g band suggests it belongs to the "LSNCC" class, similarly to the case shown in Figure \ref{fig:example-258}. Figure \ref{fig:example-691022126-flat} shows the same system at a different time. Although the four objects are more clearly visible in the image, the time series appears more flat and does not exhibit the typical peaks of exploding supernovae. There are several possible explanations for this. One hypothesis is that the supernova has already exploded and the brightness change is no longer detectable. Another possibility is that real data are inherently more variable than simulated data and noise makes peaks difficult to detect.

\begin{figure*}
 \centering
 \includegraphics[width=0.8\textwidth]{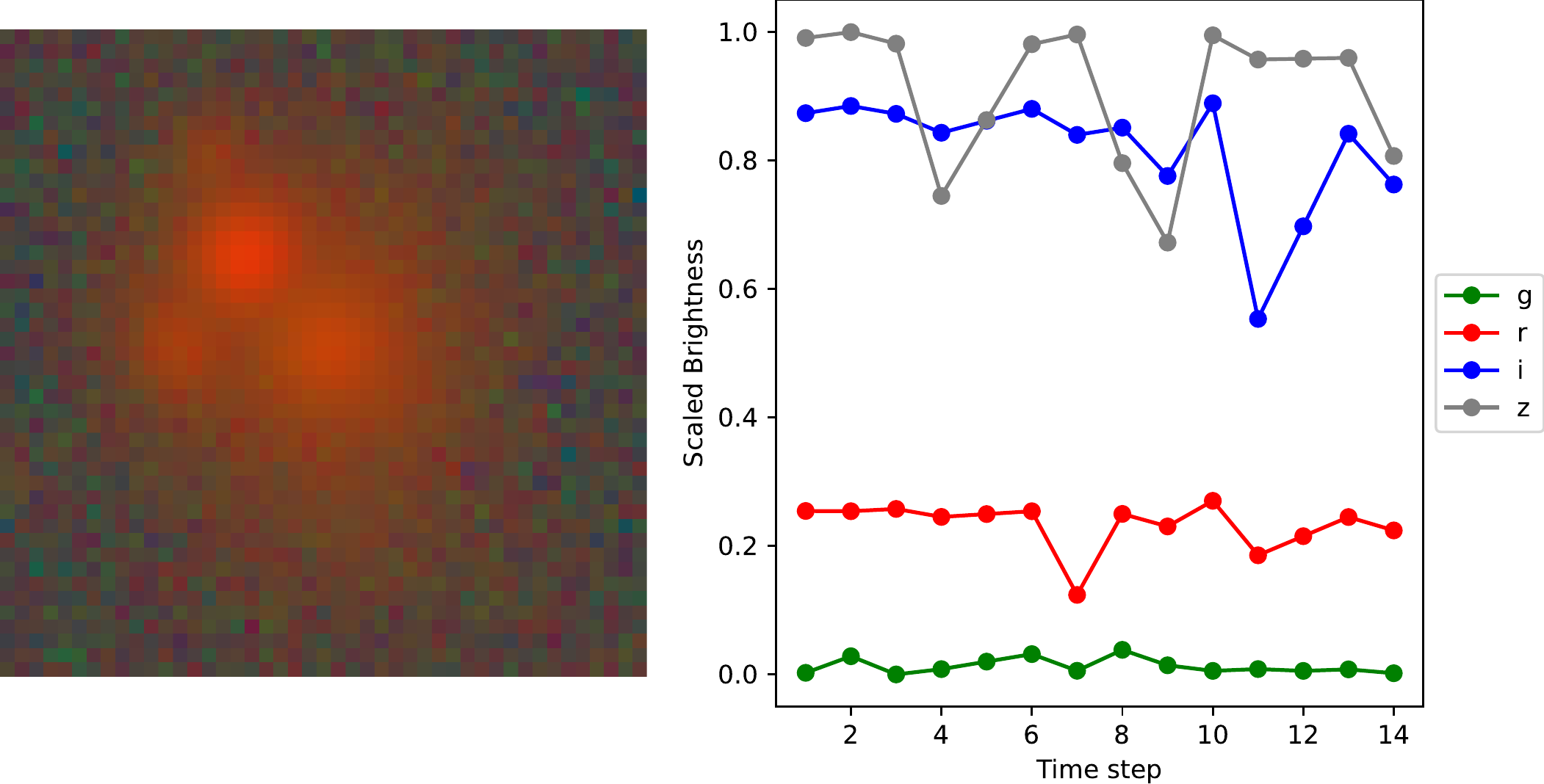}
 \caption{\textbf{A real gravitational lens} -- The system presented in this figure has been classified as a gravitationally-lensed supernova by \cite{morgan2022deepzipper2}. However, the detection was performed on the fifth year of the observation, which is not publicly available. At the time of the observation, the lens is already present, but the supernova explosion is not visible yet. The time series, indeed, are almost flat or noisy}
 \label{fig:example-699919273-flat}
\end{figure*}

Figure \ref{fig:example-699919273-flat} presents the system identified with ID 699919273, which exhibits a clear gravitational lens. Additionally, this system contains multiple objects, which are likely to be lensed versions of the same astrophysical object. The authors of  \cite{morgan2022deepzipper2} classify this system as a gravitationally-lensed supernova, based on Y5 data (not publicly available). With the available data up to Y4, the system is classified as a "Lens," which confirms the category assigned by \cite{morgan2022deepzipper2} with the public data up to Y4.

\begin{figure*}
 \centering
 \includegraphics[width=0.8\textwidth]{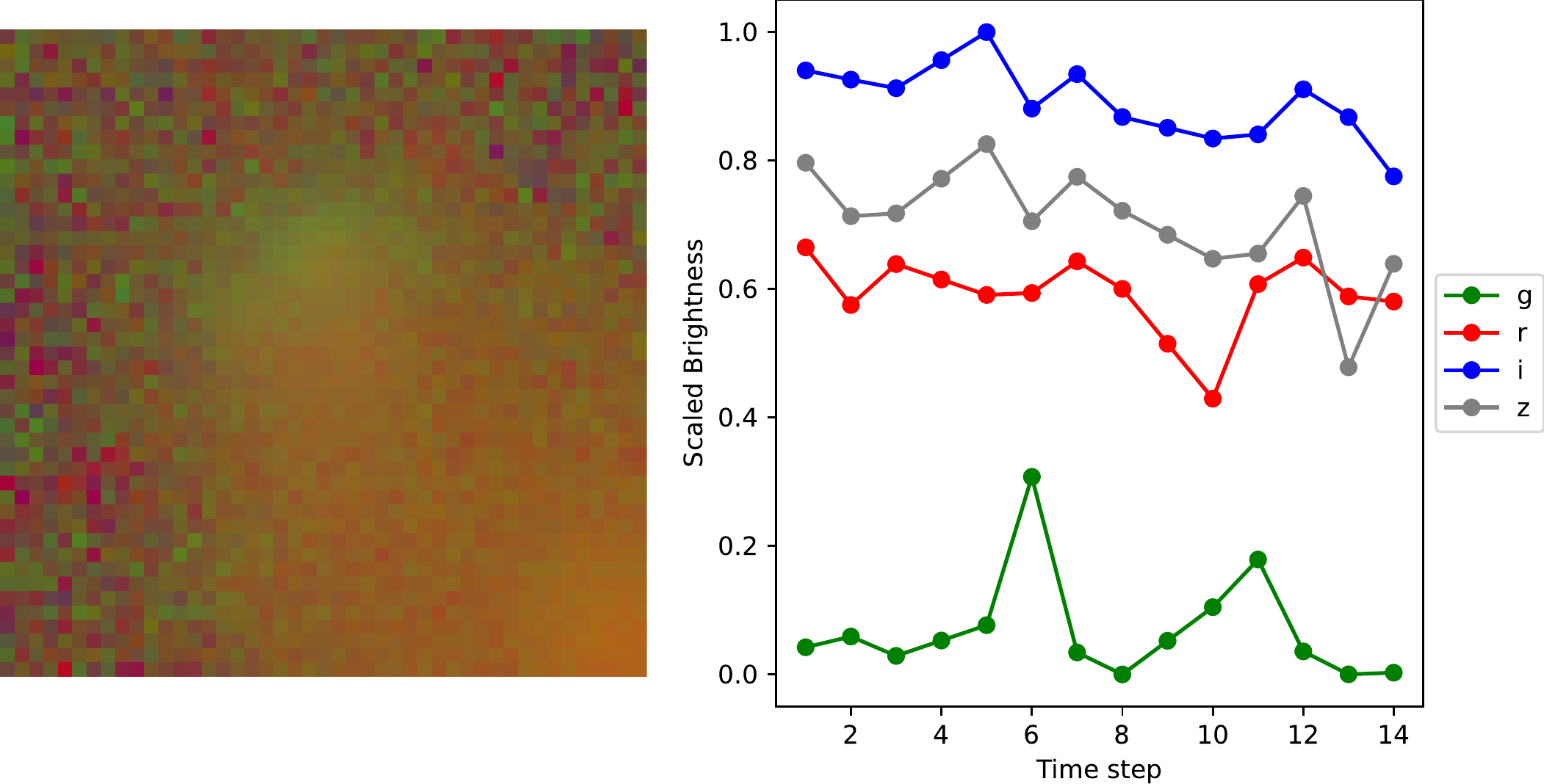}
 \caption{\textbf{A real gravitational lens} -- The system presented in this figure has been indicated as a gravitationally-lensed supernova by \cite{morgan2022deepzipper2}. However, the detection was performed on the fifth year of the observation, which is not publicly available. Before, the lens is already present, but the supernova explosion is not visible yet. The time series, indeed, are almost flat or noisy}
 \label{fig:example-701263907-flat}
\end{figure*}

Figure \ref{fig:example-701263907-flat} presents the more complex system with ID 701263907, in which the identification of individual objects is challenging due to their blurred boundaries. The presence of halos around the central bodies and in the bottom-right corner of the image suggests the existence of a gravitationally-lensed object. It is possible that the lens extends beyond the boundaries of the image, further complicating its identification. The absence of evident peaks in the time series data suggests the absence of transient phenomena. Specifically, the peaks observed in the g band do not correspond with significant peaks in other bands, indicating the absence of relevant transient effect. Similar to system 699919273, data up to Y4 hint at the presence of a lens, which \newnetworkname correctly identifies.

\section{Conclusions and Future Work}
\label{sec:conclusions}

This work has introduced \newnetworknamenospace, a  neural architecture for the classification of simulated and real gravitational lensing phenomena that processes multi-modal inputs by means of sub-networks focusing on complementary data aspects. \newnetworkname surpasses the state-of-the-art accuracy results by $\approx 3\%$ to $\approx 11\%$ on four simulated data sets with different data quality. In particular, it attains a $4.5\%$ performance increase on the LSST-wide data set, which simulates the acquisitions of the Vera C. Rubin Observatory whose operations are scheduled to start in 2023. The Vera C. Rubin Observatory is expected to detect hundreds to thousands of lensed supernovae systems, which represents a breakthrough with respect to the capacity of previous instruments. The enormous amount of data that will be acquired demands highly accurate and fast computer-aided classification tools, such as \newnetworknamenospace.

Future work will concentrate on the application of \newnetworkname to real observations as soon as they become available. The envisioned research work will also pursue the objective of creating a scientist-friendly system that allows experts to import and manually classify data from real observations to create a non-simulated data set and compute relevant classification and object detection metrics for automated data analysis, following an approach similar to the one implemented in \cite{torres2020odin, torres2021odin, zangrando2021odin}.
Finally, we plan to employ the multi-modal architecture designed for \newnetworkname for the analysis of other (possibly non-astrophysical) data sets characterized by images and time series.

\section*{Acknowledgments}
\noindent The authors thank Robert Morgan, author of \cite{morgan2022deepzipper}, for having provided guidance in the use of the \texttt{deeplenstronomy} simulator and the generation and use of the data set, and professors Hans Georg Schaathun and Ben David Normann for proofreading the initial version of the article.

\bibliographystyle{acm}
\bibliography{references}

\end{document}